 \documentclass[final,5p,times,twocolumn,authoryear]{elsarticle}

\usepackage{amssymb}
\usepackage{lipsum}
\usepackage{url} 
\usepackage{amsmath}

\usepackage{graphicx}
\usepackage{txfonts}

\usepackage{color}
\usepackage{ulem}

\usepackage{hyperref}
\hypersetup{
	colorlinks=true,        
	linkcolor=blue,         
	citecolor=cyan,         
}

\usepackage{multirow} 
\usepackage{float} 

\usepackage{natbib}

\journal{High Energy Astrophysics}

\begin{document}

\begin{frontmatter}



\title{Numerical simulations of jet launching and breakout from collapsars}


\author[label1]{Gerardo Urrutia}
\author[label1]{Agnieszka Janiuk}
\author[label2]{Hector Olivares}
\affiliation[label1]{organization={Center for Theoretical Physics, Polish Academy of Sciences},
            addressline={Al. Lotnikow 32/46}, 
            postcode={02-668}, 
            state={Warsaw},
            country={Poland}}

\affiliation[label2]{organization={Departamento de Matematica da Universidade de Aveiro and Centre for Research and Development in Mathematics and Applications (CIDMA), Campus de Santiago},
            addressline={}, 
            postcode={3810-193}, 
            state={Aveiro},
            country={Portugal}}

\begin{abstract}

Long Gamma-Ray Bursts (LGRBs) are often associated with the collapse of stripped-envelope massive stars. Powerful relativistic jets drill through the stellar envelope before the gamma emission. Previous hydrodynamical studies imposed jets artificially, neglecting accretion dynamics, while the central engine simulations have reproduced jet launching via the Blandford-Znajek mechanism focusing on the inner core regions. However, both the central engine and the progenitor structure are crucial to determining the jet's evolution. In this study, we present axisymmetric (2.5-D) GRMHD simulations that self-consistently follow jet formation from the black-hole horizon to breakout at the stellar surface ($R_\star \sim 10^{10}$~cm). The setup assumes a Kerr black hole with spin $a \sim 0.9$ in the centre of three progenitor models, varying the magnetic-field strength and geometry. Relativistic jets are successfully launched by a strong dipolar magnetic field ($B_0 \gtrsim 10^{12}$-$10^{14}$~G) from magnetically arrested disks. These jets, initially magnetically dominated, convert energy into thermal and kinetic during their propagation. We found breakout times within $1.8 \lesssim t_{\rm bo} \lesssim 3.5$~s and luminosities $L_j \sim 5\times10^{49}-7\times10^{52}$~erg\,s$^{-1}$. Our results highlight the role of the initial magnetic field strength and its geometry, emphasizing the progenitor's density distribution as a key factor impacting the final structure and dynamics of LGRB jets.

\end{abstract}




\end{frontmatter}




\section{Introduction}

Gamma-Ray Bursts (GRBs) are extremely luminous and highly variable pulses of $\gamma$-ray with luminosities of $L\sim 10^{50}-10^{52}$~erg s$^{-1}$. These bursts often last less than a hundred seconds \citep[e.g.,][]{kumar15}. They are produced from the energy dissipation in a relativistic jet \citep[e.g.,][]{paczynski91,kumar15}, whose origin is attributed to either mergers of binary neutron stars \citep[or a black hole and a neutron star, e.g.,][]{rosswogRamirezRuiz2002,RosswogRamirezRuiz2002Coales,RosswogLiebendorfer2003}, or to the collapse of stripped-envelope massive stars \citep[e.g.,][]{MacFadyenWoosley1999,WoosleyHeger_2006,Moesta2014}.

GRBs from collapsars are referred to as Long GRBs because the observed gamma radiation (the \emph{prompt emission}) lasts from tens to hundreds of seconds ($2 \lesssim t \lesssim 100$~s). In addition, they are often associated with a supernova explosion of type Ib/c \citep[e.g.,][]{Frailetal1997,perley14,Izzo2019Nature}. A powerful relativistic jet is launched from a black hole accretion disk system, called the \emph{central engine} \citep[e.g.,][]{paczynski91,Woosley1993,MacFadyenWoosley1999,Burrows_2007,Moesta2014,magnetorotational1,gottlieb2022b}. An alternative explanation proposed e.g. by \cite{Thompson1995,Qin1998,Levan2006magnetar,Metzger2008,berger2011} invokes a magnetar as the central engine powering relativistic jets in GRBs.

Numerical simulations have played a pivotal role in enhancing our understanding of how these jets are launched from the central engine. The collapsar model provides an explanation of how the cores of fast-rotating massive stars ($M_\star \sim 25 M_\odot - 30 M_\odot$) with magnetic fields of order of $B_0\sim 10^{10}-10^{12}$~G, undergo collapse into a highly spinning ($a\sim0.9$) black-hole. The magnetic flux is extracted due to \citet{bz77} mechanism, that becomes dominant source of power for  bipolar jets \citep{MacFadyenWoosley1999,MacFadyen_2001,WoosleyBloom2006,KomissarovBarkov2009,BrombergTchekhovskoy2016,bugli2021,bugli2023,Burrows2023,Vartanyan2025,Burrows2025}. In addition, the magnetorotational core collapse scenario considers a highly magnetized core ($B_0\sim 10^{14}-10^{16}$) due to the presence of a proto-neutron star (PNS). Then the magnetic flux contributes to the jet launching after the magnetically driven explosions \citep{Obergaulinger2006axisimetricMagneto,Obergaulinger2006A-2,Burrows_2007,Moesta2014,magnetorotational1,magnetorotational2,magnetorotational3,magnetorotational4,Shibata2024spinEv,FujibayashiShibata2025,Shibata2025arXiv}. Nonetheless, performing these sophisticated simulations is computationally demanding due to the use of a tabulated equation of state, neutrino transport, as well as the evolution of the mass of the black hole (BH). The integration times are focused on the jet launching and its propagation during the first fraction of a second.

Studies of long-term propagation of the jet commonly omit the accretion phase and assume a successful crossing of the iron core. Jets are initialized by imposing strong shock boundary conditions at the iron-core/outer-shell interface \citep[e.g.,][]{Begelman1989,Aloy2000-2D,Aloy2002stability,lopezcamara2009,Nagakura_2011,lopezcamara2013,lopezcamara2016,Hamidani2017,harrison18,Gottlieb2020_hydro_jets,Gottlieb2020weakly,urrutia22_3D,Suzuki2022,Pais2023collapsars,Urrutia2023Gws}. These works explore a wide range of jet luminosities, Lorentz factors, variability, structure, magnetization, and progenitor stratification in combination to evaluate their impact on jet propagation beyond breakout at large scales.

Imposing jets by hand risks erasing important dynamical effects: the accretion-launch connection influences the large-scale jet evolution dynamics on large scales. Including the early jet interaction with disk winds \citep{Murguia_Berthier_2021, urrutia2025}, the MRI-driven variability \citep{JaniukJames2022, gottlieb2022a}, or jet wobbling/precession \citep{gottlieb2022b}. The central engine in collapsars can involve complex physical scenarios. The evolution of the spin plays an important role \citep[e.g.,][]{Janiuk2023collapse,Jacquemin-Ide2024,ColemanFernandez2024outflows,Shibata2024spinEv}, along with heavy-element nucleosynthesis \citep{2022ApJ...934L..30J,ColemanFernandez2024disks,Issa2025CollpasarsNuc,Shibata2025arXiv,FujibayashiShibata2025}. However, highly magnetized environments and spinning BH facilitate efficient jet production. Therefore, the study of GRB progenitors can be limited to these key factors.

In fact, recent studies \citet{gottlieb2022a,gottlieb2022b} have followed jet production from the horizon and its propagation beyond the stellar envelope. In these works, the stellar envelope is modeled using an analytical profile based on fits to collapsed profiles from \citet{Halevi2023ApJ}, along with steeper and constant density profiles representing the inner core. While a range of slopes and magnetizations is explored, it is important to note that, in all cases, successful jets are launched with very high energy ($L \gtrsim 10^{54}$~erg~s$^{-1}$), which in some instances exceeds typical values estimated for observed GRBs ($L \sim 10^{49}$-$10^{52}$~erg~s$^{-1}$;  \cite{Nagakura2013NeutrinoJets,kumar15,Hamidani2017,Aloy2018Range,Aguilera-Dena2020,Salafia2023structuredjets,Angulo2024,enegazzi2025SNexplo}).

Focusing on highly magnetized environments and fast-spinning BHs, we perform axisymmetric General Relativistic Magnetohydrodynamic (GRMHD) simulations to study the production and propagation of jets from collapsars. The aim of this study is to investigate the effects of the stellar progenitor and magnetic field configuration on jet formation at the BH horizon and their influence on large-scale jet dynamics post breakout.

This work is organised as follows. In Section \ref{sec:methods} we describe the progenitor stars involved in this study, the initial conditions, and the configuration of the numerical setup. In Section \ref{sec:results} we present our results, and the discussion is provided in Section \ref{sec:discussion}. Finally, the conclusions are presented in Section \ref{sec:conclusions}.


\section{Methods}\label{sec:methods}

In this section, we describe our numerical scheme, physical scenario, and assumptions. Our focus is on jet launching from a pre-collapse stripped-envelope star. Accordingly, the initial conditions are characterised by imposing a structure for the GRB progenitor, including its rotation and magnetisation. We configure the setup to extract energy from a central black hole and to launch jets primarily via the Blandford–Znajek process.

\subsection{Numerical Setup}

We perform 2.5-dimensional GRMHD simulations using the Adaptive Mesh Refinement (AMR) code BHAC \citep{Porth2016,Olivares2019a} to follow the evolution of a long GRB jet, from its origin in the accretion flow to its breakout from the pre-collapse massive star. We employ a spherical logarithmically spaced grid in modified Kerr-Schild coordinates \citep[see][for more details]{Porth2016}. The computational box extends in the radial direction up to $1\times 10^5$~$r_g$, or $\sim 10^{11}$~cm by setting the BH mass $M_{\rm BH}=3M_\odot$ (or $5M_\odot$) at the gravitational radius $r_g=GM_{\rm BH}/c^2$. The azimuthal direction covers $\theta \in [0,\pi]~$rad. We use $3$ levels of refinement ($n_l=3$), and we employ $288$ cells in the $r$-direction, which increase to $1152$ at the maximum refinement level. While in the $\theta$-direction, we use $64$ cells, increasing to $248$ at the highest refinement level. The maximum integration time $t_f$ employed in our simulations is $t_f \geq 2 \times 10^5$~$t_g$, where $t_g=r_g/c$. The final time in seconds~[s] is given in Table~\ref{tab:models_performed}.

The scheme used by our simulations is a finite volume method with piecewise parabolic reconstruction (PPM) \citep{colella_piecewise_1984}, a total variation diminishing Lax-Friedrichs (TVDLF) Riemann solver, and a two-step method for time integration \citep[for more details on these and other methods available in BHAC, see][]{Porth2016}. In order to maintain the solenoidal constraint for the magnetic field, we employ the upwind constraint transport scheme by \citet{zanna_echo_2007}.

We use automated adaptive mesh refinement (AMR) based on the L\"ohner scheme \citep{lohner_adaptive_1987}, combined with the geometric criterion of enforcing refinement on the jet when it reaches the outermost parts of the simulation.

To set up the initial conditions, we use BHAC's capability to read numerical initial data. For this purpose, we prepare a data file generated from stellar evolution codes (details in Section \ref{sec:prog_star}) in the appropriate format. We follow the remapping method described in \url{https://bhac.science/tutorial-numerical-initial-data/}.

\subsection{Density and pressure of the progenitor star} \label{sec:prog_star}

\begin{figure}
\centering
\includegraphics[scale=0.3]{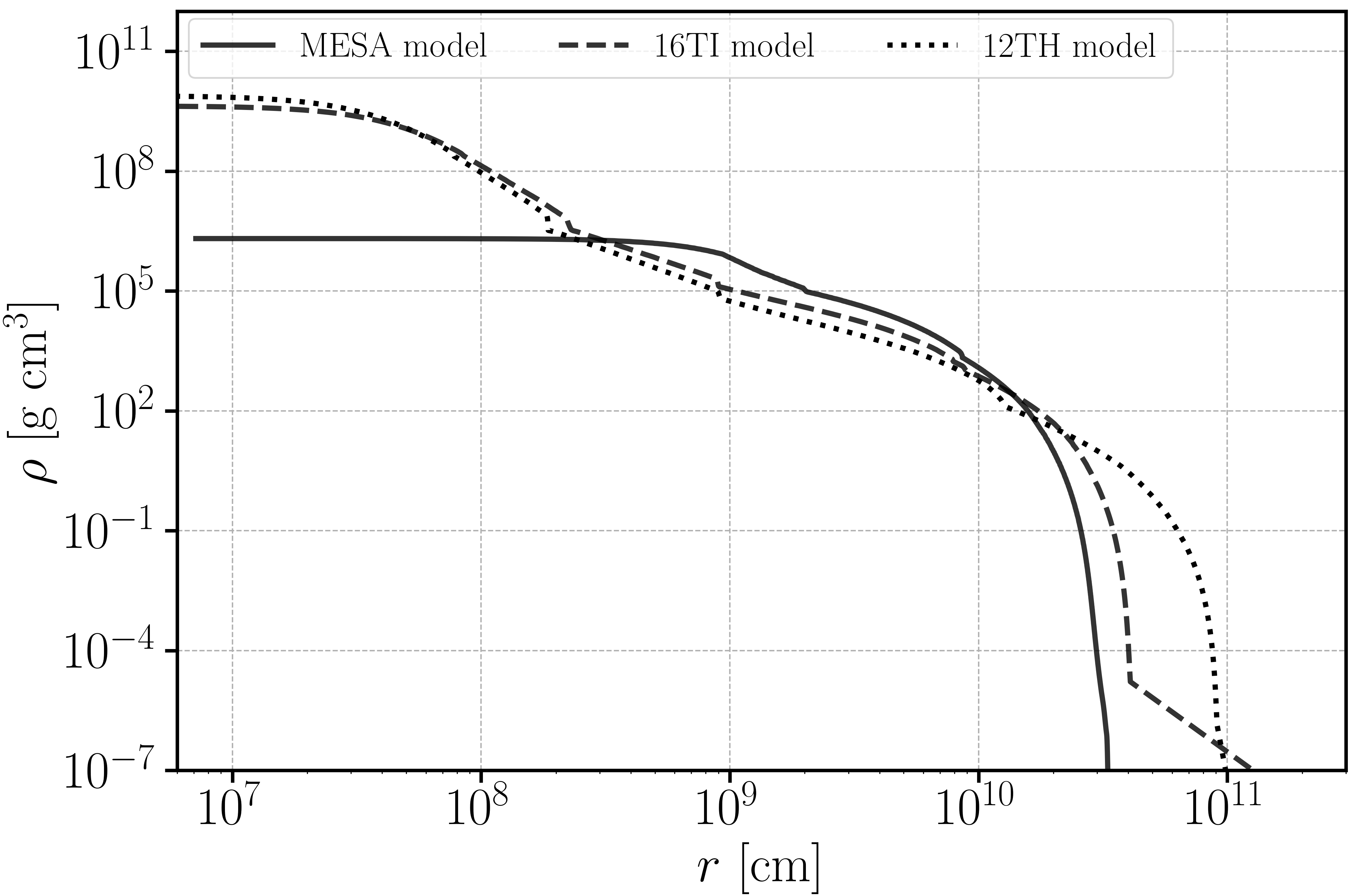}
\caption{A comparison of the density profiles of the progenitor stars. The three models are characterised by three key regions where density changes abruptly: the central region ($r\lesssim 10^8$~cm), from the boundary of the core until $r\sim 10^9$~cm, and the drop near the stellar surface.}
\label{fig:star_profile}
\end{figure}

We consider three progenitor models. The first progenitor represents the final evolutionary phase of a massive rotating star, which has evolved to form a carbon-oxygen core at its centre. The stellar structure utilised in our work was computed using MESA \citep{Paxton_2013,Paxton_2015,Paxton_2018,Paxton_2019}. The star's initial mass at the Zero-Age Main Sequence (ZAMS) was $M_{\mathrm{ZAMS}}=40\,M_{\odot}$, with a very low pre-main-sequence metallicity of $Z=10^{-5}$. After the main sequence phase, the star evolves under the assumption of a rotation frequency equal to $\Omega/\Omega_{\mathrm{crit}}=0.75$, relative to the breakup frequency. Rotation-driven mass loss is modelled using the ``Dutch'' stellar wind prescription \citep{Glebbeek2009} with a scaling factor of $f_{\mathrm{wind}}=0.8$. We obtained the density and pressure profile of the progenitor star (Figure~\ref{fig:star_profile}), which has a mass of $M_\star=25\,M_\odot$, and a stellar radius of $R_\star=3.3\times10^{10}$~cm. For this progenitor, we set a black hole mass of $M_{\mathrm{BH}}=5\,M_\odot$ at the centre.

The density and pressure profiles for the second and third progenitors were taken from \cite{WoosleyHeger_2006}, specifically the Wolf-Rayet star models 12TH and 16TI. The model 12TH has a final stellar mass of $M_\star=9.23\,M_\odot$ and radius of $R_\star=9\times10^{10}$~cm, while the model 16TI has a final stellar mass of $M_\star=13.95\,M_\odot$ and radius of $R_\star=4\times10^{10}$~cm. For both models, we set a black hole mass of $M_{\mathrm{BH}}=3\,M_\odot$. 

In the three stellar models utilised in this work, $M_{\mathrm{BH}}$ is independent of the total mass of the star $M_\star$. The black hole masses utilized in our study fall within the lower limits according to the prescription of \citet{Fryer2012}. However, detailed simulations can reveal whether collapse occurs, and determine the final BH mass \citep[e.g.,][]{magnetorotational1,Shibata2025arXiv}.

\subsubsection{Rotation}

The angular momentum distribution in progenitor stars before collapse approximately follows the form $l(r,\theta)=l(r)\sin^2(\theta)$ \citep[e.g.,][]{WoosleyHeger_2006,Burrows_2007,Lindner2010}. In our initial conditions, we adopt an analytical prescription for the rotation of the star based on this profile of the angular momentum. We set a Kerr BH with spin $a=0.9375$ for which the innermost stable circular orbit (ISCO) radius \citep[e.g.,][]{Rezzolla2016,Murguiaberthier2020,KrolJaniuk2021} is given by
\begin{equation}
    r_{\rm isco} = 3+Z_2\mp\sqrt{(3-Z_1)(3+Z_1+2Z_2)}\, ,
\end{equation}
where,
\begin{equation}
Z_1\equiv 1+(1-a^2)^{1/3}\left[(1+a)^{1/3}+(1-a)^{1/3} \right]\, ,    
\end{equation}
and
\begin{equation}
Z_2\equiv\sqrt{3a^2+Z_1^2}.    
\end{equation}
In Kerr space-time, the energy and angular momentum at $r_{\rm isco}$ are described respectively by
\begin{equation}
\epsilon_{\rm isco}=-u_{t,{\rm isco}}=\frac{1-2/r_{\rm isco}+a/r_{\rm isco}^{3/2}}{\sqrt{1-3/r_{\rm isco}+2a/r_{\rm isco}^{3/2}}}\, ,
\end{equation}
\begin{equation}
l_{\rm isco}=u_{\phi,{\rm isco}}=\frac{r_{\rm isco}^{1/2}-2a/r_{\rm isco}+a^2/r_{\rm isco}^{3/2}}{\sqrt{1-3/r_{\rm isco}+2a/r_{\rm isco}^{3/2}}}\, ,
\end{equation}
and the above values define the $\phi$-component of velocity in our initial conditions as,
\begin{equation}
u^\phi=C\sin^2\theta\left(-g^{t\phi}\epsilon_{\rm isco} + g^{\phi \phi} l_{\rm isco} \right)\, .
\end{equation}
We assume the normalization parameter $C=2$, as studied in the context of mini accretion discs in collapsars \citep{Murguiaberthier2020}.

\subsubsection{Magnetisation}

\begin{table*}[]
    \centering
    \renewcommand{\arraystretch}{1.2} 
    \begin{tabular}{|c||c|c|c|c|c|c|cc|}
    \hline
    Model & Progenitor star & $M_{BH}$  & $\rho_{\rm max}^{\star}$ & $B_{0}$ & $\sigma_{\rm max}$ & Geometry                                 & Final time & ($t_f$)  \\
    & & & [g cm$^{-3}$] & [G]  &  &                                  & [$\times 10^3~t_g$] & [s] \\ \hline \hline
    m1-B0  &  &  & & $10^{14}$ & $2.163\times 10^{-1}$ & & 200 & 5.00  \\ 
    m1-$10^{-1}$B0  & MESA & $5M_\odot$ & $2 \times 10^{6}$ & $10^{13}$ & $2.163\times 10^{-3}$ & Eq.~\eqref{eqn:mag_field} &  269 & 6.70 \\ 
    m1-$10^{-2}$B0  &&&& $10^{12}$ & $2.163\times 10^{-5}$ &  & 272 & 6.88 \\ \hline
    m2-B0   & MESA & $5M_\odot$ & $2 \times 10^{6}$  & $10^{14}$ & $2.163\times 10^{-1}$ & Eq.~\eqref{eqn:field2} & 80 & 2.06\\ \hline 
    m16TI & 16TI & $3M_\odot$  &  $4 \times 10^{9}$ & $10^{14}$ & $1.047\times 10^{-4}$ & Eq.~\eqref{eqn:mag_field}  &  354 & 5.22 \\
    m12TH & 12TH & & $8 \times 10^{9}$  &  &  $5.798\times 10^{-5}$ & & 412 & 6.07  \\ \hline
    m1-zero         & MESA & $5M_\odot$ & $2 \times 10^{6}$ & Zero   & Zero   &  None & 200 & 5.0 \\ \hline
    \end{tabular}
    \caption{
    The list of the performed models categorised by the progenitor star, the maximum density $\rho_{\rm max}^*$ at the BH horizon, the BH $M_{\rm BH}$, the magnetic field strength $B_0$ and its geometry. Additionally, the integration time for each simulation is provided in $t_g$ units and seconds.
    }
    \label{tab:models_performed}
\end{table*}

For the geometry of the magnetic field, we adopt a dipole like field \citep{Burrows_2007,Moesta2014,magnetorotational1} where the azimuthal component of the magnetic vector potential is
\begin{equation}
A_\phi = B_0\, \frac{r_c^3}{r^3 + r_c^3}\, r \sin \theta\,.
\label{eqn:mag_field}
\end{equation}
Here, the radius of the stellar core takes the value $r_c=10^{7}~$cm. We check the ratio between magnetic and kinetic energy $\sigma=B^2/8\pi\rho c^2$ to set the magnetic field strength $B_0$. Notice that the peak of magnetization is given by the maximum density $\rho_{\rm max}^\star$. As shown in Table~\ref{tab:models_performed}, the magnetic field strength considered in our models is in the range $B_0 = 10^{12} - 10^{14}$~G, that are consistent with known magnetars \citep[e.g.,][]{KaspiandBeloborodov2017}, but are lower than those found in magneto-rotational core collapse simulations whose strengths $B_0\lesssim 10^{16}$~G support the jet launching \citep{Moesta2014,magnetorotational1}. Nevertheless, as we will show below, our maximum initial magnetic field strength $10^{14}$~G should be able to produce enough magnetic flux at the BH horizon to activate the BZ mechanism, following the relation described by \citet{Burrows_2007,KomissarovBarkov2009,gottlieb2022a}, via 
\begin{equation}
  \Phi_{\rm BH,min}=4 \pi r_h^2 |B_h| \approx 7\times 10^{27}\sqrt{\frac{\rho_{\rm max}^*}{10^7~{\rm g}~{\rm cm}^{-3}}}\,{\rm G}\,{\rm cm}^{-2}\, ,
\label{eqn:mag_condition}
\end{equation}
where the subindex $_h$ refers quantities at the BH horizon.

\begin{figure*}[!h]
    \centering
    \includegraphics[width=1.0\linewidth]{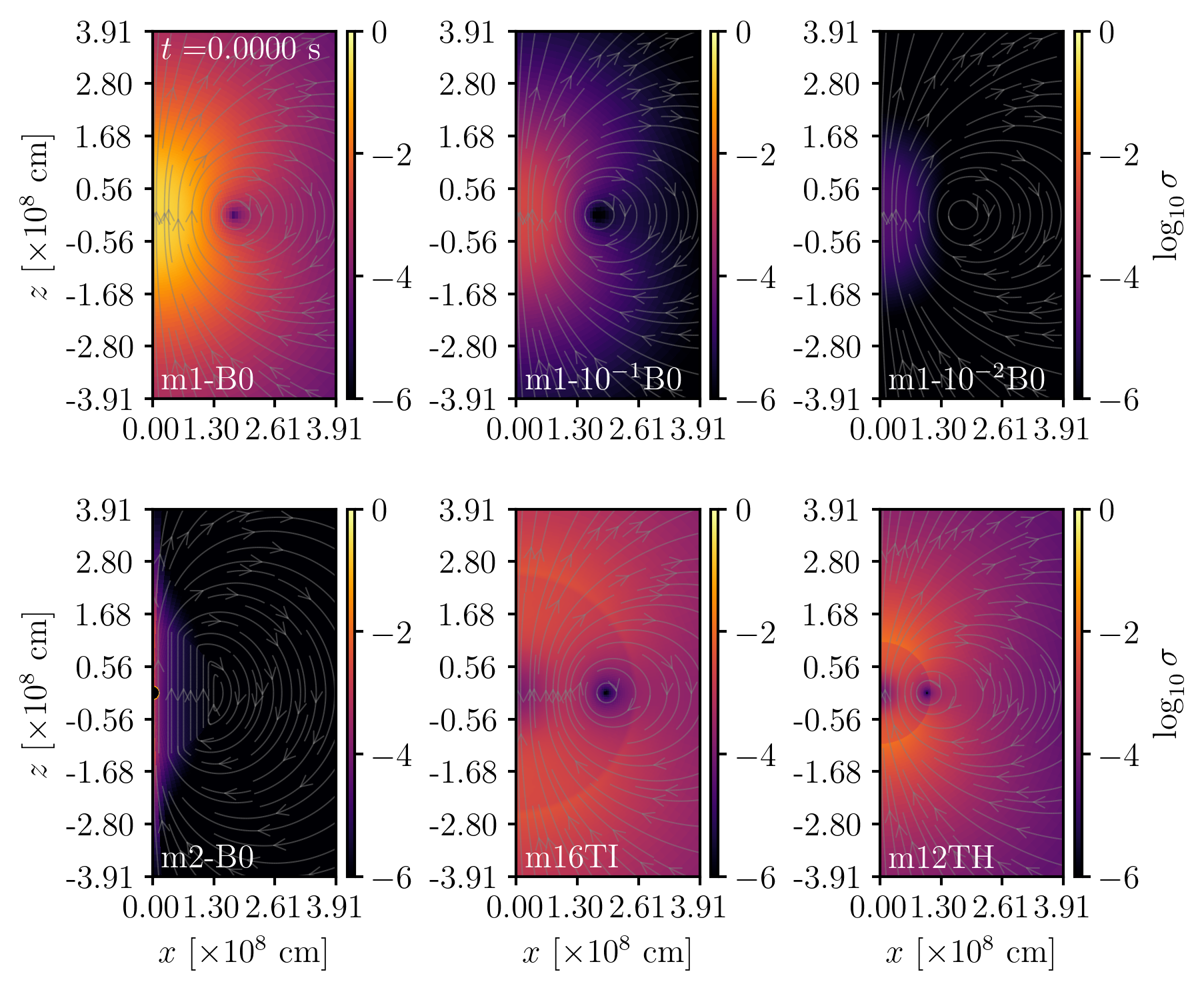}
    \caption{Initial distribution of magnetisation $\sigma$ at $t=0$. The magnetic field has a dipole-like configuration in most models. In model m2-B0 the field is vertical in the core, and outside it is dipole-like. The initial strength of the magnetic field is given in Table \ref{tab:models_performed}. }
    \label{fig:init1}
\end{figure*}

Following the magnetic field geometry utilized in \citet{gottlieb2022a}, we also adopted in one model the azimuthal component of the magnetic vector potential as
\begin{equation}
    A_\phi=B_0\,\frac{r_c^3\sin\theta}{r} {\rm max}\left(\frac{r^2}{r^3+r_c^3}-\frac{R_\star^2}{R_\star^3+r_c^3},0 \right)\, ,  
    \label{eqn:field2}
\end{equation}
were inside of the stellar core, we set a constant field $B=B_0\hat{z}$. The maximum initial $\sigma$ for our models is listed in Table~\ref{tab:models_performed}. We show the initial $\sigma$ distribution in Figure~\ref{fig:init1}, which depends on the magnetic-field geometry and both the stellar density and pressure profiles.

In the first row of Figure~\ref{fig:init1}, we show the effect of varying the magnetic field strength, $B_0 = 10^{14}, 10^{13}, 10^{12}$~G, for the same progenitor star and the same magnetic potential. The effect of changing the magnetic field configuration is shown in the first panel of the second row, while the effect of changing the progenitor star, keeping the magnetic field fixed, is shown in the second and third panels of the second row.

\section{Results}\label{sec:results}

In this section, we present the results of our simulations. We separately examine the effects of the magnetic field strength, its geometry, and the influence of stellar structure on jet dynamics. Additionally, we present various diagnostics of central engine activity as a function of time, as well as the extraction of jet luminosity.

\subsection{The effects of the magnetic field strength and geometry}\label{sec:results_magnetic_field_strenght}

\begin{figure*}[!h]
    \centering
   \includegraphics[width=1.0\linewidth]{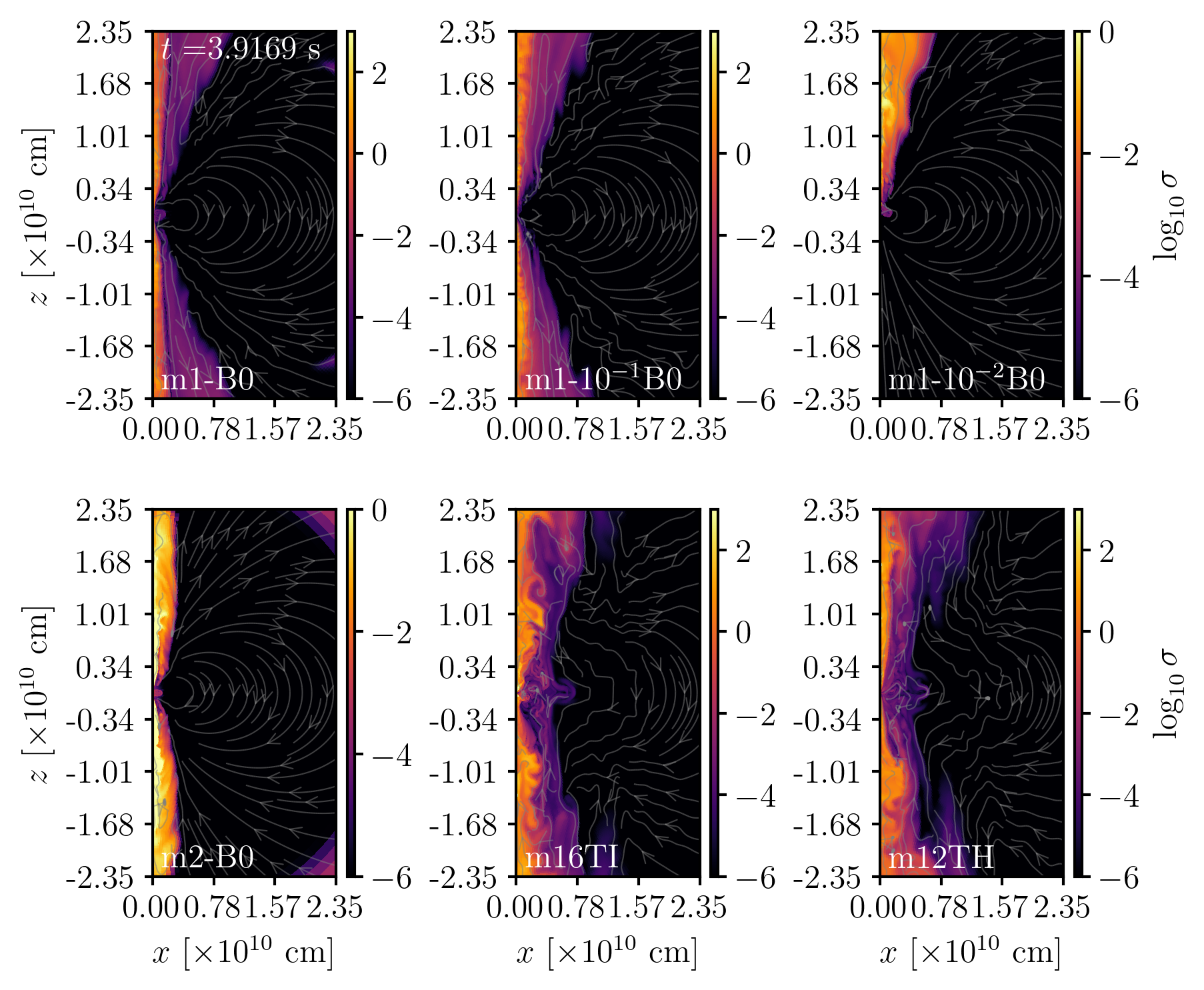}
     \caption{The map of the magnetisation and magnetic field lines at $t\sim 3.9$~s. We define the jet core as the high magnetised material $\sigma \geq 1$, while the jet wings the region where $10^0>\sigma > 10^{-6}$.}
   \label{fig:init2}
\end{figure*}

As the magnetised core material falls and begins accreting onto the black hole, the jet is launched by the magnetic flux generated via the Blandford–Znajek mechanism. A disk-like structure is formed, MRI transports magnetic energy, and the magnetic flux in the centre increases, then the magnetised outflow drives a shock front that sweeps material outward from the denser inner layers, as is shown for axisymmetric cases in \citet{Burrows_2007,Moesta2014}.

In Figure~\ref{fig:init2}, we present maps of the the jet magnetisation. Here we define the jet core as the region where $\sigma \geq 1$, while the jet wings have $10^0>\sigma > 10^{-6}$. We observe the impact of the initial magnetic field strength on the jet dynamics in the models m1-B0, m1-$10^{-1}$B0, and m1-$10^{-2}$B0, which were commonly initialized with the MESA progenitor and the magnetic potential given by Eq.~\eqref{eqn:mag_field}. The models m1-B0 and m1-$10^{-1}$B0 clearly show a collimated region $\sigma>1$ in the jet axis, and jet wings that are approximately two orders of magnitude less magnetised, allowing two jet components to be readily distinguished. In model m1-$10^{-2}$B0, the jet exhibits a single conical outflow with a uniform magnetisation of $\sigma \lesssim 1$. As a result, the core and wings are not clearly distinguished, and the flow appears as a broad, single-component jet rather than a distinct two-component structure. In addition, this model shows the formation of a non-bipolar jet as previously shown, for example by \citet{Powell2024}.

In the first panel of the second row in Figure~\ref{fig:init2}, we present the magnetization of the model m2-B0. It was initialised combining a uniform parallel field within the stellar core with a dipole configuration outside it, as described by Eq.~\eqref{eqn:field2}. The jet got a cylindrical shape, and its core finished less magnetized with respect to the models initialised with an initial dipole.

Finally, the distribution of $\sigma$ for models m16TI and m12TH is presented in the second and third panels in the second row of Figure~\ref{fig:init2}. The jet core and the wings are regions of mixing. It is because these stars present different density jumps along their inner density structure. Then, the driven outflows transit between different shells, which induce perturbations and mixing between the core and the jet wings.

\subsection{The structure evolution of the progenitor star}

\begin{figure}[!h]
    \centering
    \includegraphics[width=1.0\linewidth]{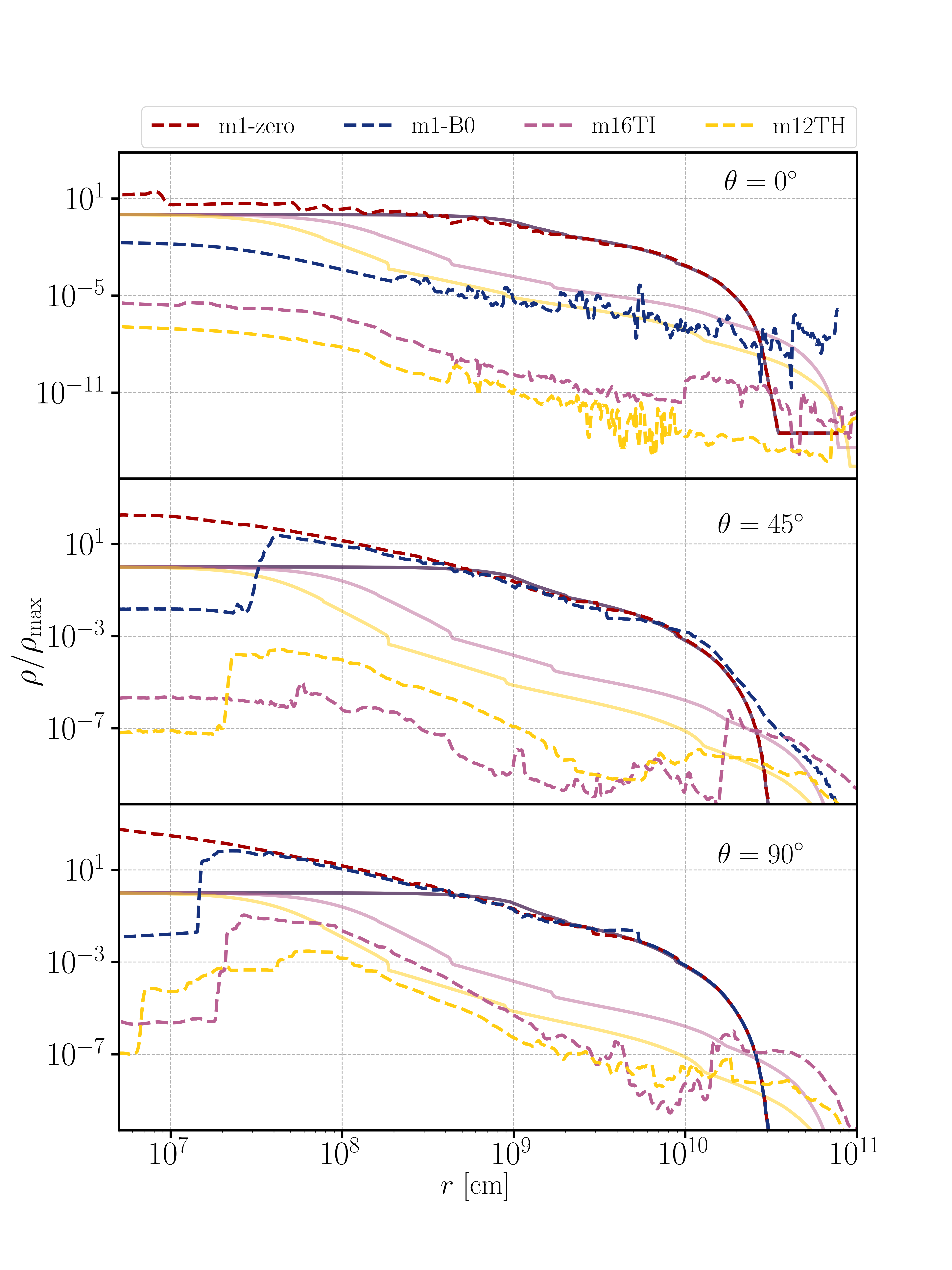}
    \caption{Radial profiles of density after five seconds of evolution. Each panel displays slices taken at different polar angles. The top panel corresponds to the $z$-axis ($\theta=0^\circ$), while the middle and bottom panels show cuts at $\theta=45^\circ$ and $\theta=90^\circ$, respectively. Continuous lines show the initial condition and dashed lines show the stratification at $t=5$~s. 
    }
    \label{fig:density_rad}
\end{figure}

Figure~\ref{fig:density_rad} shows the progenitor density structure for three GRB progenitors: MESA model, 16TI, and 12TH. Radial profiles are normalized with respect to the maximum density value. We present a comparison between distributions at $t = 0$~s (solid lines) and $t = 5$~s (dashed lines) along three polar angles: $\theta = 0^\circ$, $45^\circ$, and $90^\circ$.

The model m1-zero, where no jet is launched, at $\theta=0^\circ$ and during the evolution of the density increases by about one order of magnitude with respect to the initial value at distances $r\lesssim 10^6$~cm. At $\theta=90^\circ$, it increases by nearly two orders of magnitude. We do not find a low-density funnel produced during the evolution. However, a dense accretion disk is created.

The evolution of models m1-B0, m16TI, and m12TH, which share the same magnetic field geometry and initial strength ($B_0 = 10^{14}$~G), exhibits distinct final density distributions along the jet funnel ($\theta = 0^\circ$). The densities in models m16TI and m12TH decrease by five orders of magnitude with respect to the maximum density of model m1-B0.

At $\theta = 45^\circ$, the density distribution changes due to the presence of the cocoon, which is more extended in models m16TI and m12TH. A similar behavior is observed at $\theta = 90^\circ$ where the cocoon and disk winds affect the density profile. It drops across the entire region for models m16TI and m12TH, while model m1-B0 remains perturbed only near the central region.

\subsection{Central engine diagnostics}\label{sec:CE-diag}
\begin{figure}[!h]
    \centering
    \includegraphics[width=1.0\linewidth]{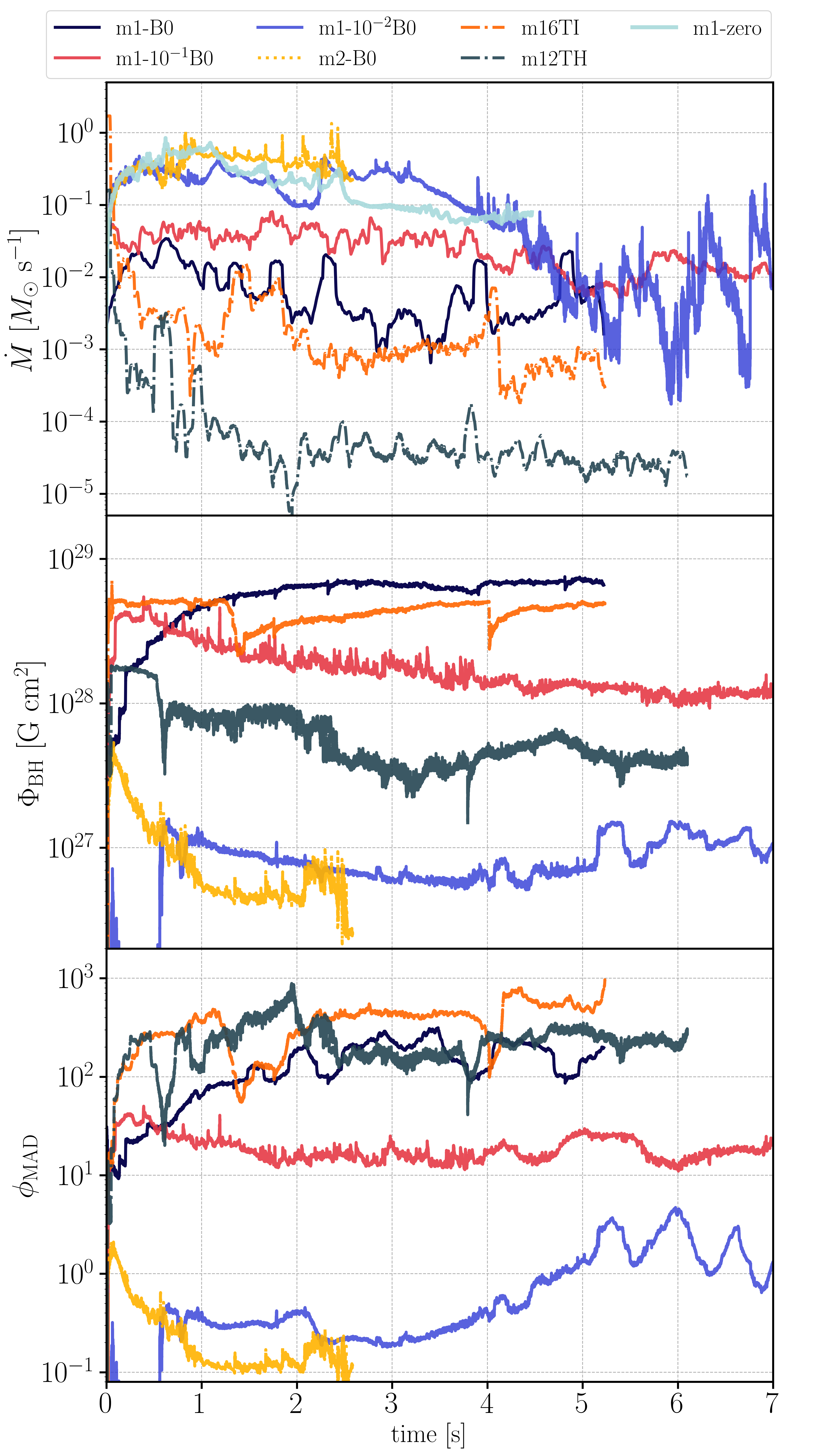}
    \caption{The central engine activity for all models. The top panel shows the accretion rate, the middle panel presents the magnetic flux, and the bottom panel displays the MAD parameter in Lorentz-Heaviside units.}
    \label{fig:ceAct}
\end{figure}
In Figure~\ref{fig:ceAct}, we present the time evolution of the central engine activity through the accretion rate $\dot{M}$, the magnetic flux $\Phi_{\rm BH}$, and the MAD parameter $\phi_{\rm MAD}$. The accretion rate is estimated by
\begin{equation}
    \dot{M}=-\iint \rho c^2 u^r \sqrt{-g} d\theta d\phi \; ,
\end{equation}
where $\rho$ is the density, $u^r$ the radial contravariant four-velocity and $g=|g_{\mu\nu}|$ is the determinant of the metric. The integration is performed at radial distance $r_{50}\equiv 50r_g$ to avoid contamination by the magnetic ceiling close to the horizon \citep[e.g.,][]{Lalakos2024,Lalakos2025}.

In the upper panel of Figure~\ref{fig:ceAct}, we show the time evolution of the accretion rate. Weakly magnetized (m1-$10^{-2}$B0 and m2-B0) and non-magnetised models (m1-zero) exhibit values in the range of $10^{-1}\lesssim \dot{M}< 1$~$M_{\odot}$~s$^{-1}$, while those of strongly magnetised jets (m1-B0, m1-$10^{-1}$B0, 16TI, 12TH) are in a range of $10^{-4}\lesssim \dot{M} < 10^{-1}$~$M_{\odot}$~s$^{-1}$. It is evident that $\dot{M}$ is modified by the ejection of the strong jets and winds whose velocities are $\Gamma u>1$. The accretion rate depends on the initial magnetic field strength $B_0$, which regulates the amount of available magnetized gas for accretion (Figure~\ref{fig:init1}). High magnetized accretion disks induce magneto-centrifugal acceleration, resulting in stronger winds and jets \citep[e.g.,][]{JaniukJames2022}.

The magnetic flux presented in the middle panel of Figure~\ref{fig:ceAct} is estimated as
\begin{equation}
    \Phi_{\rm BH}=\frac{1}{2} \iint |B^{r}| \sqrt{-g} d\theta d\phi \; ,
    \label{eqn:magnetic_flux}
\end{equation}
where $B^r$ is the radial component of the magnetic field. The highly magnetized models m1-B0, m1-$10^{-1}$B0, m16TI, and m12TH reach maximum magnetic flux values of $10^{28} \lesssim \Phi_{\rm BH} \lesssim 10^{29}$~G~cm$^2$, consistent with the condition in Eq.~\eqref{eqn:mag_condition} required to drill through the infalling material of the progenitor stars used in this study. The evolution of the magnetic flux decays slowly over the first five seconds, decreasing by less than one order of magnitude for models m1-$10^{-1}$B0, m1-$10^{-2}$B0, and m12TH. In contrast, model m2-B0 shows a rapid decline in the magnetic flux because its initial magnetized region does not extend over a large volume (Figure~\ref{fig:init1}). On the other hand, models m1-B0 and 16TH, which start with a more extended magnetized region, maintain a roughly constant saturated magnetic flux for up to six seconds. We also estimate the MAD parameter to evaluate the degree of magnetic saturation and the dominance of magnetic flux in the system.

The accumulation of magnetic flux in the accretion disk drives the system into a magnetically arrested disk (MAD) state, where magnetic pressure dominates and centrifugal forces launch outflows, as reported for the case of collapsars by \citet{gottlieb2022a,Issa2025b}. In the bottom panel in Figure~\ref{fig:ceAct}, we show the MAD parameter defined as
\begin{equation}
    \phi_{\rm MAD} = \frac{\Phi_{\rm BH}}{\sqrt{\langle \dot{M} \rangle_{\tau}r_{50}^2 c}} \; .
\end{equation}
Here, the accretion rate is averaged over a temporal window of $\tau=5000 t_g$. The MAD state is achieved in models with magnetic flux $\Phi_{\rm BH} \gtrsim 10^{27}$~G~cm$^2$, corresponding to $\phi_{\rm MAD} \gtrsim 15$ in Heaviside-Lorentz units. In these cases, jets efficiently extract rotational energy from the black hole and convert it into magnetic energy, maintaining a nearly constant $\phi_{\rm BH}$ after $t \approx 2$~s. Models 16TI and 12TH reach saturation rapidly, while m1-B0 saturates more gradually to $\phi_{\rm MAD} \gtrsim 100$, ensuring successful jet launching. In contrast, m1-$10^{-1}$B0 does not reach full saturation, and m1-$10^{-2}$B0 remains near the MAD threshold but still produces a jet due to the persistent magnetic flux contributing to the ram pressure. Model m2-B0 shows an initial peak in magnetization that triggers a jet, but the magnetic flux decays quickly, leaving the jet too weak to penetrate the stellar envelope. Less magnetized models never exceed $\phi_{\rm MAD} \sim 10$.

\subsubsection{Luminosity and variability}
\begin{figure}[!h]
    \centering
    \includegraphics[width=1.0\linewidth]{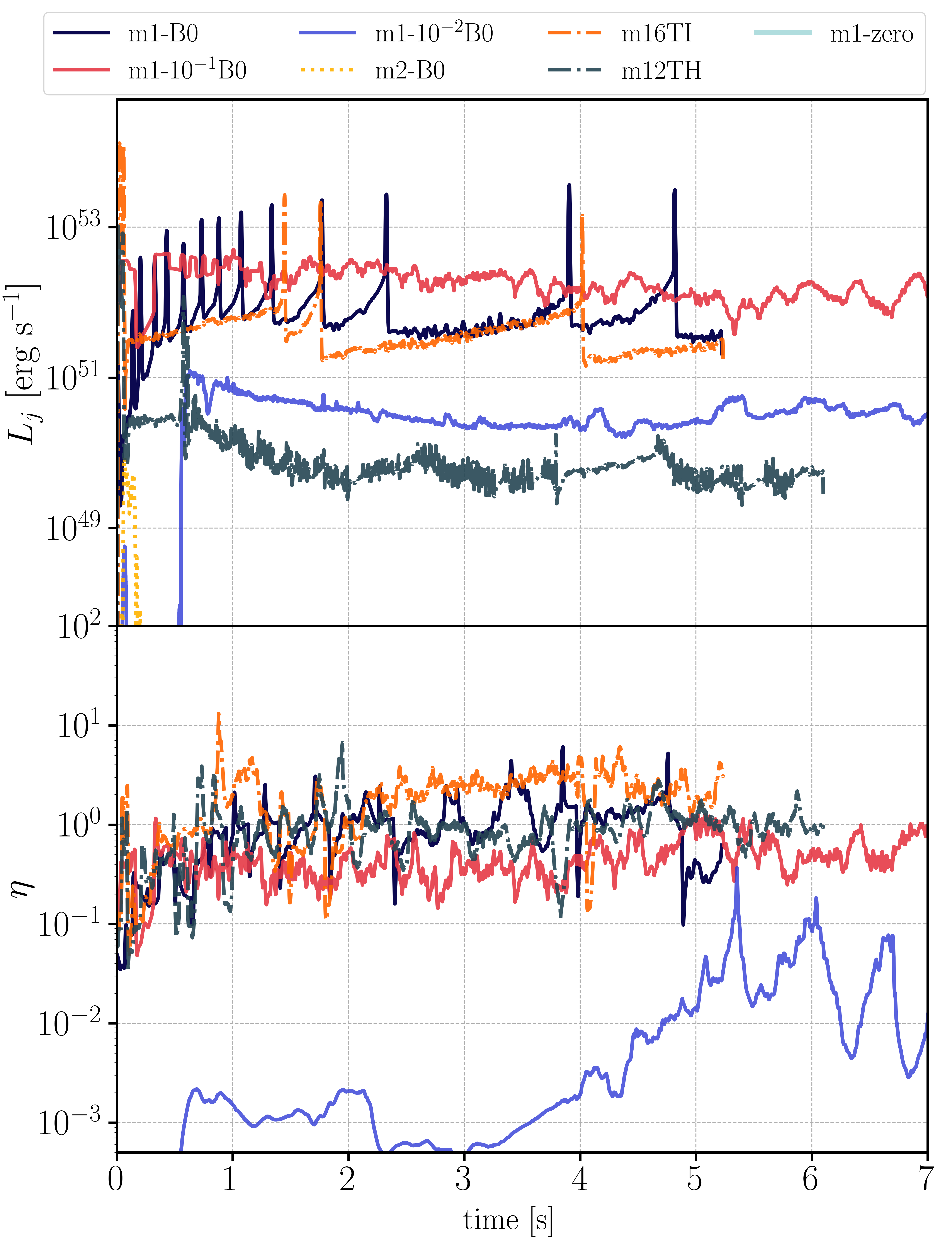}
    \caption{Jet luminosity (top panel) and the jet efficiency (lower panel). Models are labelled above the plot. The magnitude of jet efficiency can determine whether a jet is successful or a failed jet. }
    \label{fig:lum_effi}
\end{figure}

We extract the total north-pole luminosity of the jet $L_j$, considering only material with Lorentz factor $\Gamma \geq 2$ in order to distinguish jet material from winds. Three components are included: magnetic $L_{\rm EM}$, thermal $L_{\rm TH}$, and kinetic $L_{\rm KE}$. The total jet luminosity is then given by the sum
\begin{equation}
    L_j = L_{\rm EM} + L_{\rm TH} + L_{\rm KE} \; .
    \label{eqn:lum_tot}
\end{equation}
The magnetic component is extracted as
\begin{equation}
    L_{\rm EM} = -\frac{1}{4\pi} \int_{0}^{\pi/2} \int_{0}^{2\pi} \left(b^2 u^r u_t - b^r b_t \right) \sqrt{-g} \, d\phi \, d\theta \;,
\end{equation}
where $b^2=b^\mu b_\mu$, being $b^\mu$ the comoving four-component of the contravariant component of the magnetic field in the fluid frame and $u^{\mu}$ the four-components of the velocity.

The thermal component is given by 
\begin{equation}
    L_{\rm TH} = - \int_{0}^{\pi/2} \int_{0}^{2\pi} \left(u_g + p_g\right) u^r u_t \sqrt{-g} \, d\phi \, d\theta \;,
\end{equation}
where $u_g$ and $p_g$ are the thermal energy and gas pressure, and the kinetic component of the luminosity is given by
\begin{equation}
    L_{\rm KE} = - \int_{0}^{\pi/2} \int_{0}^{2\pi} \rho c^2 u^r u_t \sqrt{-g} \, d\phi \, d\theta \;.
\end{equation}
In the top panel of Figure~\ref{fig:lum_effi}, we show the time evolution of $L_j$ integrated at $r_{50}$. The profile reveals central engine variability and follows a temporal decay of one order of magnitude in a temporal window of $t\sim 7$~s.

Models m1-B0 and m16TI exhibit recurrent peaks, although the intervals are not strictly periodic. They occur more frequently during the first two seconds of evolution and become less regular at later times. These peaks arise when fluid parcels fall back and re-energize the central engine, and are further enhanced by magnetorotational (MRI) instabilities \citep[e.g.,][]{janiuk_variability2021}. Such peaks correspond to the moments of the magnetic saturation (e.g., magnetic flux in Figure~\ref{fig:ceAct}). We also analyse the jet efficiency
\begin{equation}
    \eta = \frac{L_j}{\dot{M}c^2} \;.
\end{equation}
This is shown in the bottom panel of Figure~\ref{fig:lum_effi}. The saturated models m1-B0 and m16TI reach efficiencies of $\eta > 10^{0}$, indicating that magnetic energy is extracted via the Blandford–Znajek process, i.e., the jet luminosity exceeds the energy rate of the infalling material. The other successful jets achieve efficiencies of $\eta \gtrsim 10^{-1}$ according with \citet{McKinney2005,Salafia2021AGiacomazzo2021}, while the failed cases remain in the regime $\eta < 10^{-1}$.  Such low efficiencies are consistent with jets that do not reach the MAD state, due to weaker magnetization, which limits the ability to extract enough rotational energy from the central object.

\subsection{The influence of the central engine on the jet properties beyond the star's surface}

\begin{figure}[!h]
    \centering
    \includegraphics[width=1.0\linewidth]{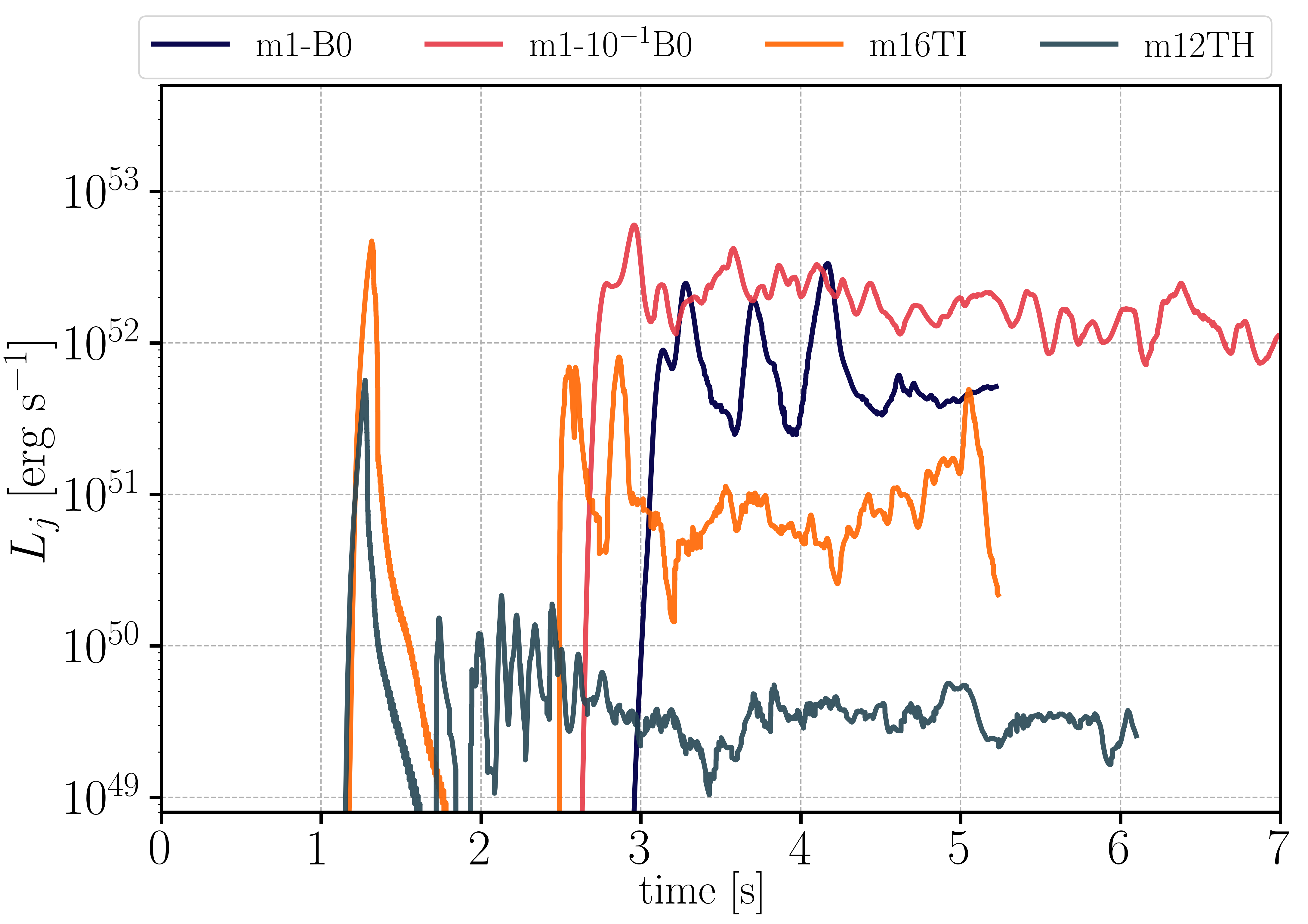}
    \includegraphics[width=1.0\linewidth]{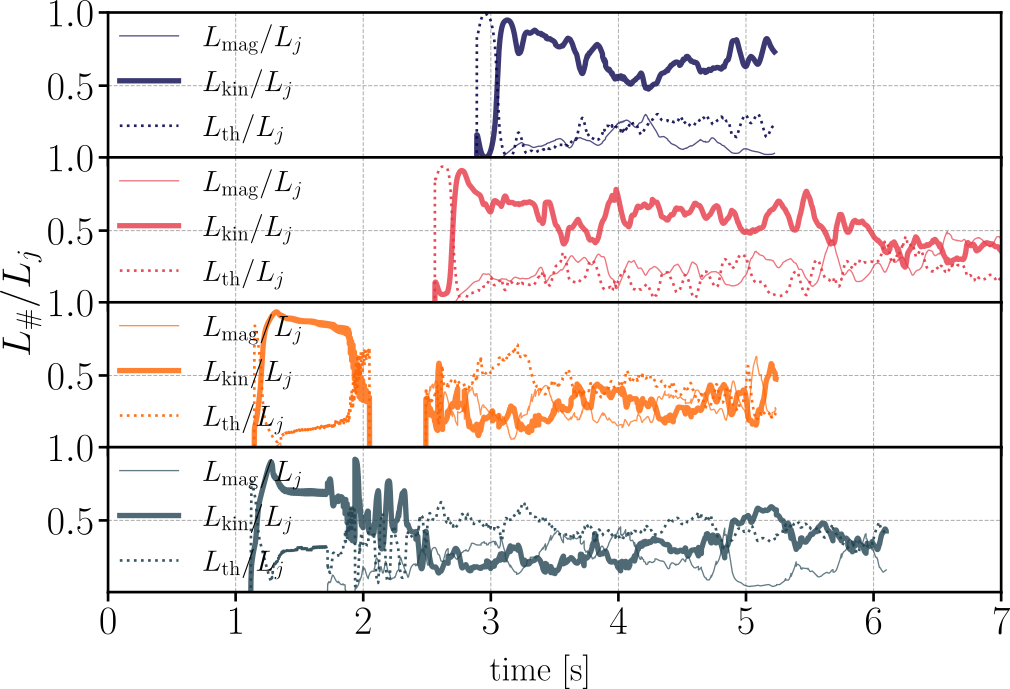}
    \caption{Upper panel: Estimation of the jet luminosity at $r = 10^{10}\,\mathrm{cm}$. Middle panel: Time evolution of the three luminosity components; magnetic ($L_{\rm mag}$), kinetic ($L_{\rm kin}$), and thermal ($L_{\rm th}$). They are normalized by the total luminosity $L_{\rm tot} = L_{\rm mag} + L_{\rm kin} + L_{\rm th}$. 
    }
    \label{fig:fraction_and_radial_energy}
\end{figure}

In Figure~\ref{fig:fraction_and_radial_energy}, the upper panel shows the jet luminosity $L_j$ estimated at the stellar surface $r=R_\star$ of the successful jets m1-B0, m1-$10^{-1}$B0, m16TI, and m12TH, and for these models. The models m1-B0 and m16TI exhibit continuous luminosity peaks at $r=r_{50}$ (Figure~\ref{fig:ceAct}), and those peaks appear at $r=R_\star$ with a time delay according to the breakout time.

\begin{figure}[!h]
    \centering
    \includegraphics[width=1.0\linewidth]{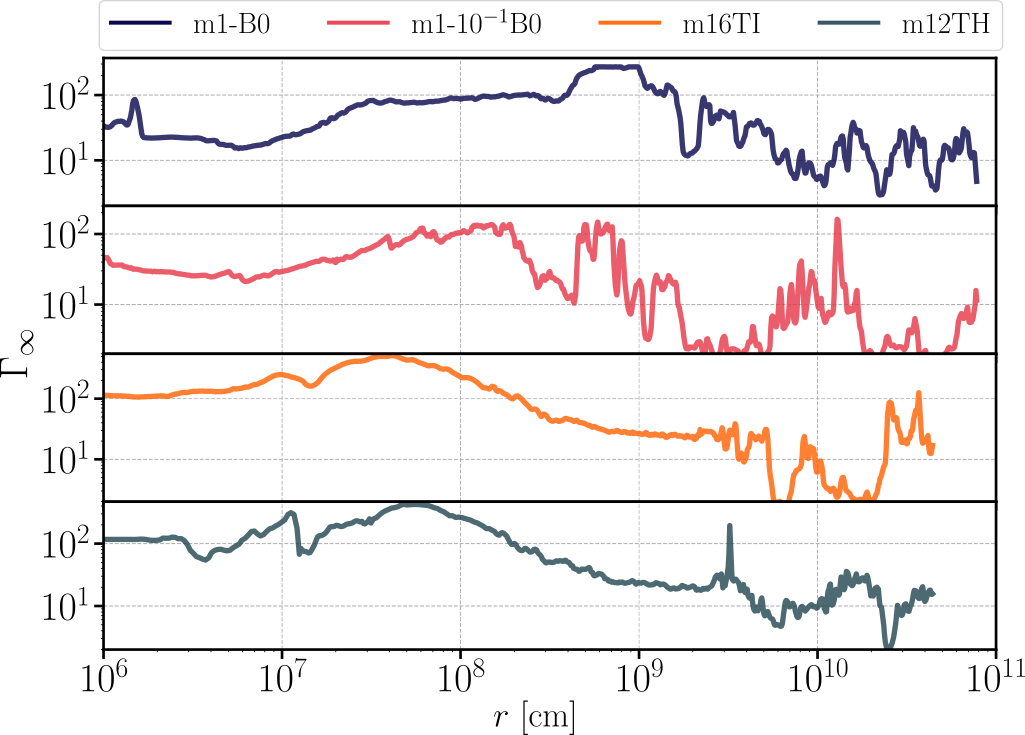}
    \includegraphics[width=1.0\linewidth]{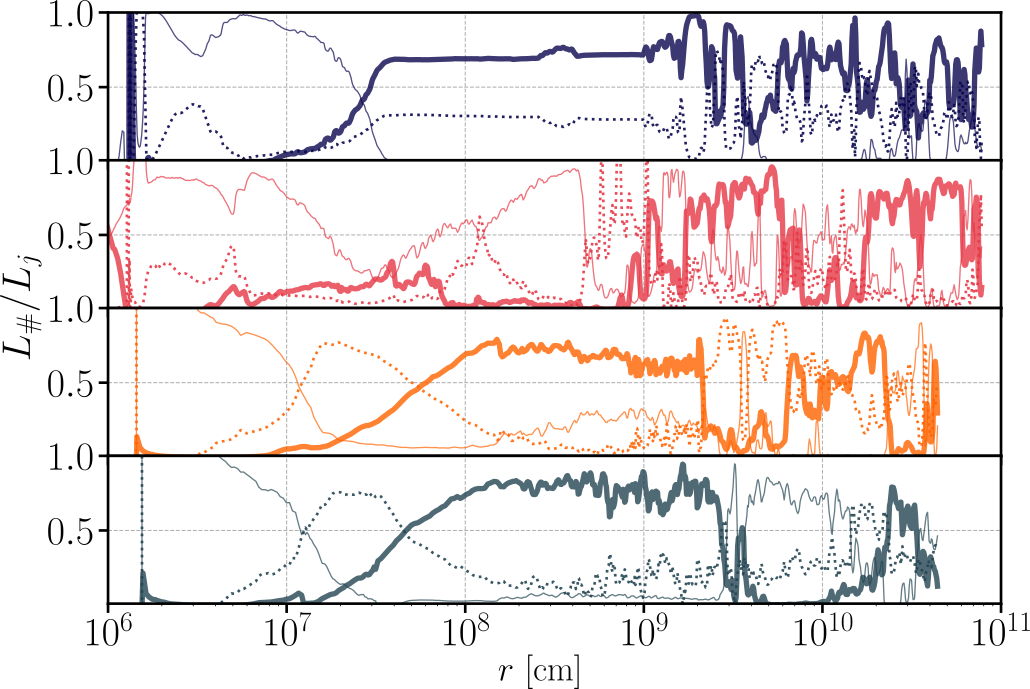}
    \caption{The radial profile of the asymptotic Lorentz factor $\Gamma_\infty$. Each row corresponds to a different successful jet model. The maximum values are consistent with regions where the conversion of magnetic energy into kinetic energy is efficient. Bottom panel: Radial stratification of the energy components in the final simulation snapshot.}
    \label{fig:lor_infty}
\end{figure}

We note in the upper panel of Figure~\ref{fig:fraction_and_radial_energy} that models m16TI and m12TH exhibit a pronounced initial peak (or flare) at the breakout time. Immediately afterward, the luminosity decreases by nearly two orders of magnitude within the first second of evolution. In MAD-state systems, similar precursor flares have been reported by \citet{GottliebUnifiedPictur2023}. Our models m16TI and m12TH display the same feature, associated with the rapid transition of the system into the MAD state (see Figure~\ref{fig:ceAct}).

The luminosities of our successful jets are in the range of $5 \times 10^{49}$ to $7 \times 10^{52}~{\rm erg~s^{-1}}$. Compared to previous 2.5 dimensional simulations, \citet{Burrows_2007} launched jets from from $M_\star = 15\,M_\odot$ stellar envelopes,  
using dipole magnetic fields whose strengths are $B_0 \sim 10^{10}-10^{12}$~G, the jets acquire a luminosity of $L_j \sim 10^{50}-10^{51}$~erg~$s^{-1}$. Similarly, \citet{KomissarovBarkov2009} produced MAD jets with $L_{\rm j} \sim a^2 \dot{M} c^2 \sim 10^{50}~{\rm erg~s^{-1}}\,(a/0.1)(\dot{M}/10^{-2})$. Our results agree well with these findings.

In the central panel of Figure~\ref{fig:fraction_and_radial_energy}, we show the temporal evolution of the energy fractions $L_\#/L_j$ for each component, with $L_\# = L_{\rm mag}, L_{\rm kin}, L_{\rm th}$. Models m1-B0 and m1-$10^{-1}$B0 display a peak in the thermal component at breakout, after which the kinetic energy quickly dominates with $L_{\rm kin}/L_j \gtrsim 0.5$. In contrast, models m16TI and m12TH show an initial kinetic peak reaching $L_{\rm kin}/L_j \sim 0.9$ at $t \sim 1$-$2$~s, followed by thermal dominance between $t \sim 2$-$6$~s. After $\sim 5$~s, the kinetic and thermal fractions approach similar values. Overall, in models m1-B0 and m1-$10^{-1}$B0, the kinetic fraction dominates over the magnetic and thermal components, whereas in m16TI and m12TH the kinetic fractions generally remain below or close to 0.5.

In the upper panel of Figure \ref{fig:lor_infty}, we show the radial profile of $\Gamma_\infty \equiv -u_t (h+\sigma)$. When this value reaches its maximum ($\Gamma_\infty \gtrsim 100$), it indicates the conversion of the jet's magnetic energy into kinetic energy \citep[e.g.,][]{gottlieb2022b}.

To examine in detail which component dominates,  in the lower panel of Figure~\ref{fig:lor_infty}, we show the luminosity fractions as a function of radius $r$, analysed from the last simulation snapshot. The magnetic component dominates in the inner region ($r < 10^7$ cm) but drops between $r \sim 10^7$–$10^8$ cm. In model m1-B0, magnetic energy is converted into kinetic energy, consistent with the results of \citet{Tchekhovskoy2009}. Model m1-$10^{-1}$B0 behaves differently: when the magnetization drops, the thermal component dominates up to $r \sim 10^9$ cm. Beyond $r \gtrsim 10^9$~cm the kinetic energy dominates, except near $r \sim R_\star$. For models m16TI and m12TH, the energy conversion follows a similar pattern: magnetization drops around $r \sim 10^7$ cm, the thermal component rises and then decreases, while the kinetic component grows and dominates with $L_{\rm kin}/L_{j}\gtrsim 0.5$. Close and beyond $r \sim R_\star$, the energy components oscillate, which implies that outer shells of the star impact the jet dynamics. This highlights the role of the progenitor environment in shaping the energy conversion.

Because the jet undergoes transitions between magnetic and kinetic energy dominance, we measure the jet opening angle $\theta_{\rm j,50}$ and $\theta_{\rm j,R_\star}$ both near the black hole horizon $r_{50}$ and at the stellar surface $R_\star$. Close to the horizon, where magnetic energy dominates, the jet region is defined by $\sigma > 1$. At breakout, we also identify ultra-relativistic material with $\Gamma u > 2$, which confirms efficient conversion of magnetic to kinetic energy. The resulting values are listed in Table~\ref{tab:final_outcomes}. We find that the jet opening angle does not have the same value when measured near the black hole horizon as it does at the scale of the stellar surface. We do not observe recollimation in $\theta_{\rm j, R_\star}$ \citep[e.g.,][]{Hamidani2021-expanding} when using $\Gamma u > 2$. However, when using the criterion $\sigma > 1$, we observe collimation. Our results suggest that after breakout, the jet core has two components enclosed in a small angle: magnetic and kinetic, showing together that the recollimation is not really happening.

\begin{table}
    \centering
    \begin{tabular}{|l|c|cc|c|}
      \hline
      Model   & $\theta_{{\rm j},50}$~[$^\circ$] &  $\theta_{{\rm j},R_\star}$~[$^\circ$] &  & $t_{\rm bo}$~[s]   \\
      & {\tiny $\sigma>1$} & {\tiny $\sigma>1$} & {\tiny $\Gamma u>2$} &    \\ \hline \hline
       m1-B0          & $8.6$ & $4.5$ & $8.4$ & $2.3$   \\
       m1-$10^{-1}$B0 & $6.4$ & $6.6$ & $8.7$ & $1.9$  \\
       m1-$10^{-2}$B0 & $6.7$ & $2.6$ & $5.2$ & $3.0$  \\
       m16TI           & $4.9$ & $8.4$ & $10.6$& $1.8$  \\   
       m12TH           & $5.8$ & $4.6$ & $7.5$ & $3.5$  \\ \hline
    \end{tabular}
    \caption{The final jet parameters. The jet opening angle $\theta_j$ was measured at two radii, $r=50$ and $r=R_\star$. We evaluate the opening angle using two criteria, based on magnetisation $\sigma > 1$ and a velocity criterion $\Gamma u > 2$, respectively, in each column. Additionally, we report the breakout time $t_{\rm bo}$.}
    \label{tab:final_outcomes}
\end{table}

\begin{figure}[!h]
    \centering
    \includegraphics[width=0.98\linewidth]{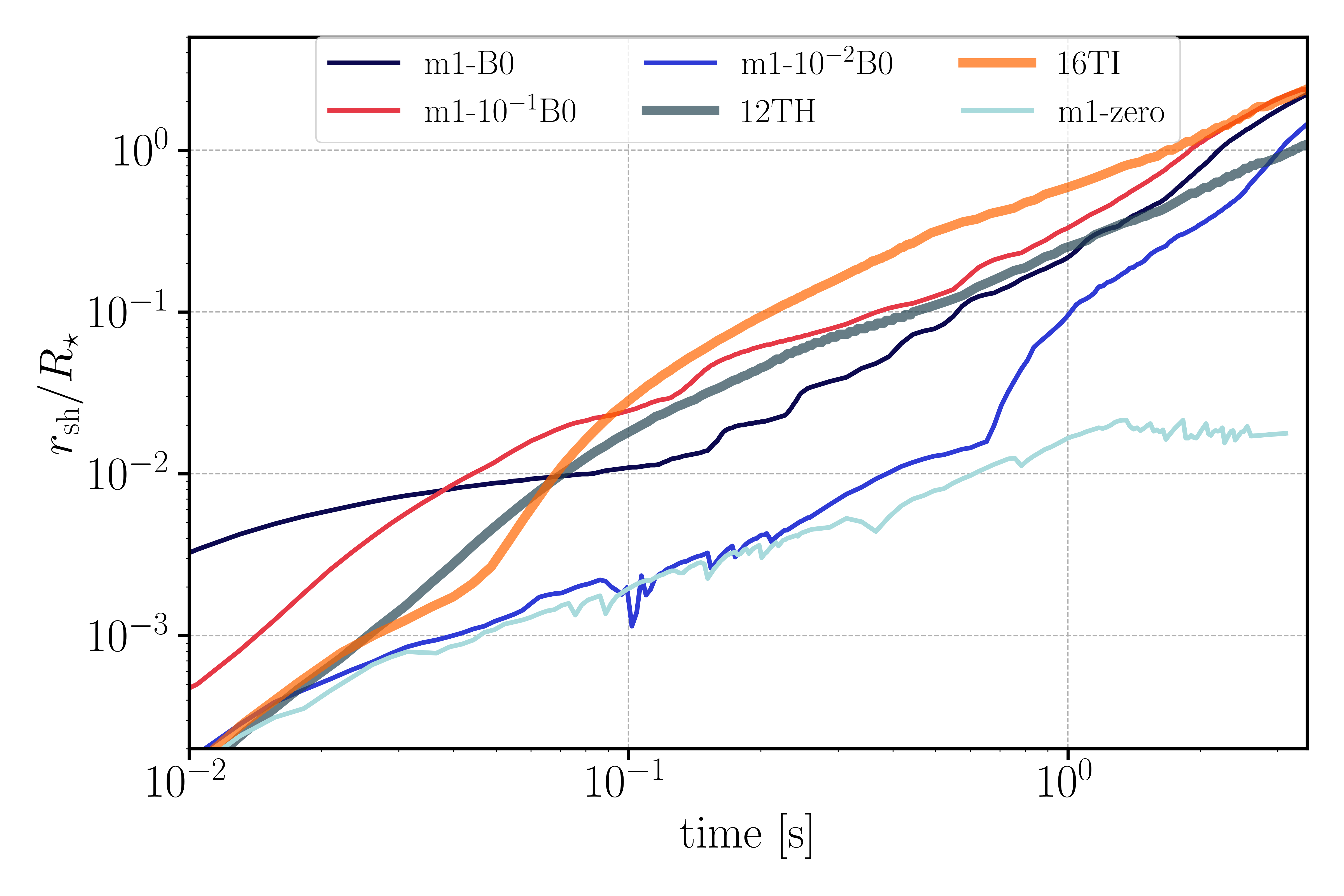}
    \caption{The evolution of the shock front normalized to the stellar radius. The breakout time is defined when the shock front crosses the progenitor boundary at $r_{\rm sh}/R_\star=1$.}
    \label{fig:shock_front}
\end{figure}

In Figure~\ref{fig:shock_front}, we show the evolution of the shock front propagation. The fastest jets (m16TI and m12TH) remain collimated near the base and break out earlier, probably it is also associated to the flare shown in Figure~\ref{fig:fraction_and_radial_energy}. In addition, this kinetic energy component dominates during expansion in the star's outer layers ($r>10^{8}$~cm). Such kinetically dominated jets undergo fast propagation, as shown in \citet{Matsumoto2019,urrutia22_3D}. In addition, we show the shock front evolution of our non-magnetized model, m1-zero. The shock front stops at approximately $t \approx 2$~s due to the absence of a Poynting flux (see SASI explosions in \citet{gottlieb2022a}) or a thermal energy source \citep[e.g.,][]{Crosato2024}. A sufficiently strong neutrino flux could revive the propagation of the shock front, as has been shown, for example by \citet{magnetorotational1}.

\begin{figure*}[!h]
    \centering
    \includegraphics[width=1.0\linewidth]{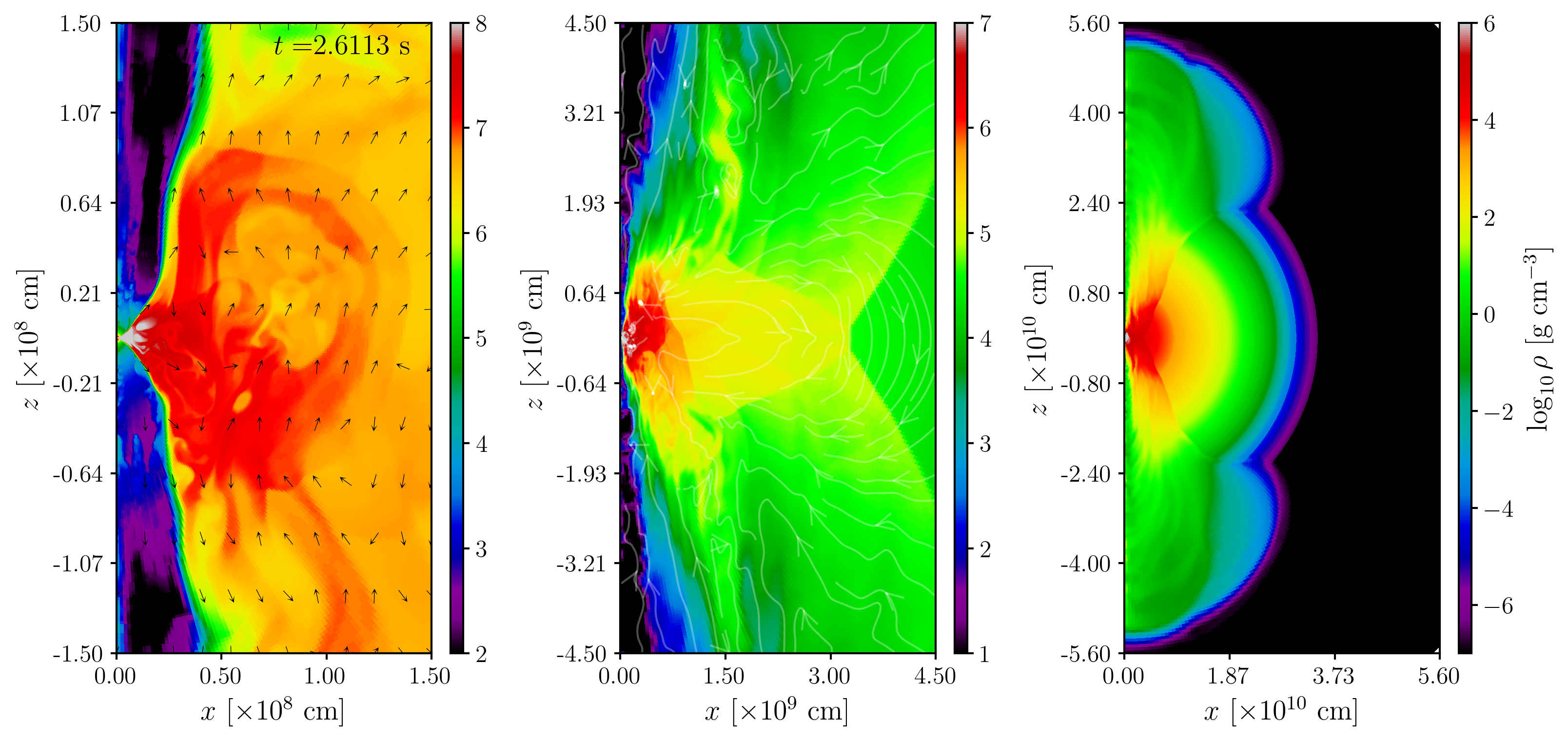}
    \caption{Three views of the same snapshot taken at time $t = 2.61$~s for model m1-$10^{-1}$-B0, zoomed in different scales: $1.5\times 10^{8}$ cm (left), $4.5\times 10^{9}$ cm (middle), and $5.6\times 10^{10}$ cm (right). The left panel focuses on the stellar core, highlighting the central engine activity with arrows indicating material being accreted or ejected. The central panel covers a region spanning $15\%$ of the total stellar radius ($R_\star$), illustrating the jet funnel and magnetic field lines. Finally, the right panel provides a full view of the stellar size, showing the jet expansion beyond its breakout.}
    \label{fig:e_model1_1d-1B0}    
\end{figure*}

Finally, we present a density map in Figure~\ref{fig:e_model1_1d-1B0} with three panels showing different zoomed-in views of our computational domain. We illustrate here the model m1-$10^{-1}$-B0 with the inner region containing the accretion disk, an intermediate region beyond the core, and the outer region beyond the stellar surface. The purpose of this map is to illustrate the jet originating from the central engine and, finally its cocoon as it engulfs the star.

\section{Discussion}\label{sec:discussion}

In the simulations presented above, we investigated the effects of magnetisation, magnetic field geometry, and the progenitor structure on the formation and dynamics of the Long GRB jet.

The simulations were initialised with three different spherically symmetric stellar progenitors, assuming a Kerr black hole had already formed. Rotation was introduced via constant angular momentum scaled to that at the innermost stable circular orbit, and either a dipolar or a hybrid magnetic field was imposed. Our simulations follow the jet propagation from the black hole horizon out to a radial distance of $10^{11}$~cm.

Below, we first compare self-consistent jet-launching mechanisms originating from the black hole with scenarios where jets are artificially imposed at the iron core surface. We then place our results in the context of previous studies. Finally, we assess the limitations of our initial conditions and assumptions, and discuss potential implications for the observational signatures of GRB progenitors. While our model does not aim to provide a fully realistic or definitive scenario, it offers valuable insights that may help bridge theoretical predictions and observations.

\subsection{Central engines in collapsars and jet launching}

Collapsars have been studied through detailed simulations, primarily focusing on supernova explosion mechanisms driven by neutrino energy deposition \cite[e.g.,][]{Burrows_2007,lopezcamara2009,Obergaulinger2006axisimetricMagneto,Lindner2010,Lindner_2012,Mezzacappa2020,Shibata2024spinEv,Jacquemin-Ide2024,Dean2021,ColemanFernandez2024disks}. While neutrinos play a crucial role up to black hole formation, their influence diminishes post-collapse, then neutrino emission is not expected to dominate in jet dynamics \citep[e.g.,][]{MacFadyenWoosley1999,Burrows_2007}. Post-collapse, the efficiency of neutrino emission rapidly drops, allowing us to assume their effects are negligible. Moreover, a highly magnetized environment can induce a magnetically driven explosion in the presence of a rapidly spinning black hole \citep[e.g.,][]{Moesta2014}. This supports the validity of our progenitor environment for jet launching studies.

The accretion process in the model m1-zero shows an increment of the central density by one order of magnitude compared to the initial conditions (see Figure~\ref{fig:density_rad}). The non-magnetised simulation (and model m2-B0) does not produce a collimated outflow. It shows that a highly magnetised jet is a necessary condition to drill through the dense material. The successful jets were produced with a BH spin $a=0.9$, and a dipole magnetic field whose strengths were $B_0 \gtrsim 10^{12}$~G. They generated a magnetic flux of $\Phi_{\rm BH} \gtrsim 10^{25}$~G~cm$^2$ according with Eq.~\eqref{eqn:mag_condition}. It is the minimum expected to power a magnetic luminosity satisfying $L_{\rm EM} \gtrsim \dot{M}_{\rm in}c^2$, thus exceeding the kinetic luminosity of the infalling material \citep[e.g.,][]{Burrows_2007,KomissarovBarkov2009,gottlieb2022a}.

The jet opening angle must be small to produce slim bow shocks in dense progenitor environments \citep[e.g.,][]{Bromberg2011,harrison18,Hamidani2021-expanding,urrutia22_3D,Urrutia2023Gws,Pais2023collapsars}. In Table~\ref{tab:final_outcomes}, we list the values of $\theta_j$ obtained using the magnetisation criterion ($\sigma > 1$), both near the horizon and at the stellar surface. Our results show $\theta_j < 10^\circ$, consistent with observed jet opening angles for long GRBs with luminosities of $L_j \lesssim 10^{52}$~erg~s$^{-1}$ \citep{Lloyd-Ronning2020}. At the stellar surface, the jet opening angle decreases when using the magnetic criterion. However, adopting a velocity-based criterion ($\Gamma u > 2$) results in a larger opening angle. This behaviour arises because some models convert part of their magnetic energy into kinetic energy. Consequently, the jet opening angle is not preserved when applying the same criterion outside the star. In this region, relativistic material provides a more reliable measure of the jet opening angle. In some cases, the jet opening angle inside the star differs from that outside the star under both criteria, indicating that the opening angle is not always preserved. This suggests that the angle measured outside the star by observations may not directly correspond to the jet opening angle at the central engine.

The propagation of our successful jets (Figure~\ref{fig:shock_front}) shows that the jets break out of the star in times shorter than those typically reported in the range $t \sim 5$–$10$~s \citep[e.g.,][]{lopezcamara2009,lopezcamara2016,Hamidani2017,harrison18,Hamidani2021-expanding,Suzuki2022,urrutia22_3D,Urrutia2023Gws,Pais2023collapsars}. This is due to the high magnetisation of our models. An improvement in the structure of the collapsar progenitor could further reduce the breakout time. Additionally, three-dimensional simulations would be required to confirm this result.

Our 2.5-dimensional simulations capture the essential global parameters and necessary conditions for jet launching. However, future work should include full 3D simulations for direct comparison. Three-dimensional models \citep[e.g.,][]{Moesta2014,BrombergTchekhovskoy2016,gottlieb2022a,magnetorotational1} produce more asymmetric structures and turbulence-driven magnetic energy transfer. Another effect observed in 3D simulations is jet wobbling \citep[][]{gottlieb2022b}.

In our simulations, the high spin parameter $a = 0.9$ guarantees sufficient magnetic flux to collimate jets with a dipolar magnetic field \citep[e.g.,][]{MacFadyenWoosley1999,Burrows_2007,BrombergTchekhovskoy2016,gottlieb2022a}. We obtain jet luminosities of $10^{49}\lesssim L \lesssim 10^{53}\,\mathrm{erg\,s^{-1}}$, consistent with typical observed GRB values \citep[e.g.,][]{Salafia2023structuredjets,Angulo2024}. Simulations with lower, constant spins \citep[e.g.,][]{Gottlieb2023lowspining,Issa2025b} produce jets with luminosities $L_j \lesssim 10^{52}\,\mathrm{erg\,s^{-1}}$, an order of magnitude lower than our most magnetised case. In collapsar models, simulations have shown that the spin can decrease to $a \approx 0.2$ within approximately one second \citep[e.g.,][]{Janiuk2023collapse}, and analytical estimates suggest that the evolution of the luminosity $L_j(t)$ can be modified by longer timescales \citep[e.g.,][]{Jacquemin-Ide2024}. In addition, more complex structures of the magnetic field can regulate the luminosity of the jet, producing less energetic successful jets \citep{bugli2021,bugli2023}.

Regarding progenitor models, we also adopt the 16TI profile, which is commonly used in previous GRB simulations \citep[e.g.,][]{MacFadyenWoosley1999,Aloy2000-2D,lopezcamara2013,lopezcamara2016,Hamidani2017,Urrutia2023Gws}. However, fully self-consistent stellar evolution simulations often find that this model does not undergo direct collapse to a black hole \citep[e.g.,][]{lopezcamara2009,Nagakura_2011,magnetorotational1}. Instead, a magnetar may form and launch a jet \citep[e.g.,][]{Moesta2014,magnetorotational1,magnetorotational2,magnetorotational3,magnetorotational4}. Therefore, one must take care when assuming an existing Kerr black hole within this progenitor profile or when attributing jets to be launched by BHs. A self-consistent check of black hole formation is necessary for setups that assume a Kerr BH.

Progenitor models used in numerical simulations of jet launching are often implemented through analytical density profiles \citep[e.g.,][]{gottlieb2022a,gottlieb2022b,Issa_2025,Issa2025b}. These profiles typically follow a radial dependence $\rho(r) \propto r^{-\alpha_p}(1 - r/R_\star)^\delta$, where the exponents $\alpha_p$ and $\delta$ approximate the structure of the progenitor environment during collapse. This approach was motivated by \citet{Halevi2023ApJ}, who simulated the first milliseconds of collapse for the GRB progenitor models of \citet{Aguilera-Dena2020} and derived corresponding density fits. In particular, the core region was characterized by $\alpha_p = -1.5$.

Such analytical implementations have shown an useful technical control of the initial conditions and link the time dependence of outflows with the structure of the progenitor. For instance, \citet{Issa2025b} reported a dependence between the core density slope and the accretion rate. Models with $\alpha_p = 0$ yielded $\dot{M} \propto t^{-0.06}$, while $\alpha_p = -1.5$ produced $\dot{M} \propto t^{-1.3}$. When compared with our simulations, the case $\alpha_p = 0$ roughly corresponds to the core of our MESA progenitor. Our results are into the limit $\dot{M}\propto t^{-5/3}$.

While analytical profiles can approximate stellar collapse, assuming a uniform structure fails to capture the layered composition of realistic progenitors. In this work, we instead start from pre-collapse MESA progenitor and models 16TI/12TH to study the conditions, in very dense GRB-like progenitors, required for jet launching and to test the viability of our setup for massive stars. In MHD simulations, preserving realistic density and magnetization distributions is crucial, as our results show that the outer layers of the progenitor significantly influence jet propagation. Density discontinuities in these layers induce perturbations in the jet, affecting its stability and evolution. In addition, our results agree with luminosity and breakout times of the expected MAD jets estimated from pre-collapsed long GRB progenitors reported by \citet{Morales-Rivera2025}.

Finally, we study the strong winds produced by the non-magnetised progenitor. For example, \citet{Murguiaberthier2020} followed the expansion of the shock front over hundreds of gravitational times in the case of a low-spinning BH. In our work, we increase the spin to guarantee sustained strong winds over longer durations, allowing us to track the shock front (Figure~\ref{fig:shock_front}) and explore whether it can eventually break out of the star. At a late time, however, the shock front stops its expansion. This is due to the fact that there is no thermal energy source, for example, neutrino heating. It demonstrates that the disk wind in a non-magnetised system is not enough to break the star; it needs a thermal energy source \citep[e.g.,][]{Crosato2024,Crosato2025}.

\begin{figure}[!h]
    \centering
    \includegraphics[width=0.9\linewidth]{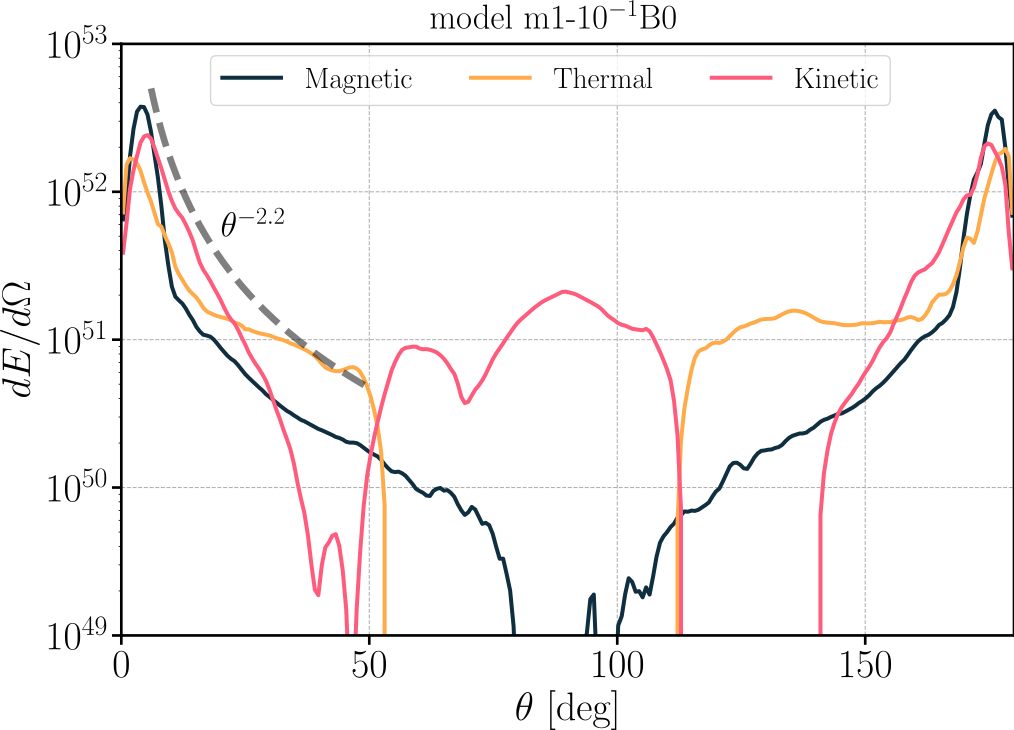}
    \includegraphics[width=0.9\linewidth]{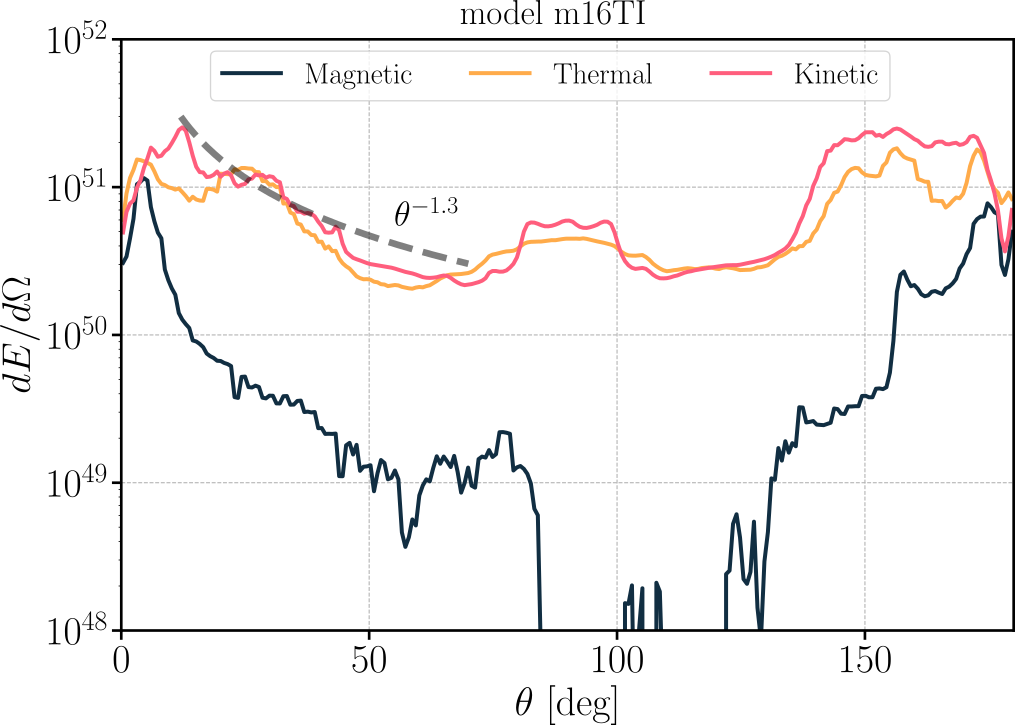}
    \caption{The final energy structure. The upper panel shows the results taken from model m1-$10^{-1}$B0, while the lower panel corresponds to model m16TI. The dashed gray lines show the fit of the slope.}
    \label{fig:energy_final_structurse}
\end{figure}

\begin{figure}[!h]
    \centering
    \includegraphics[width=0.9\linewidth]{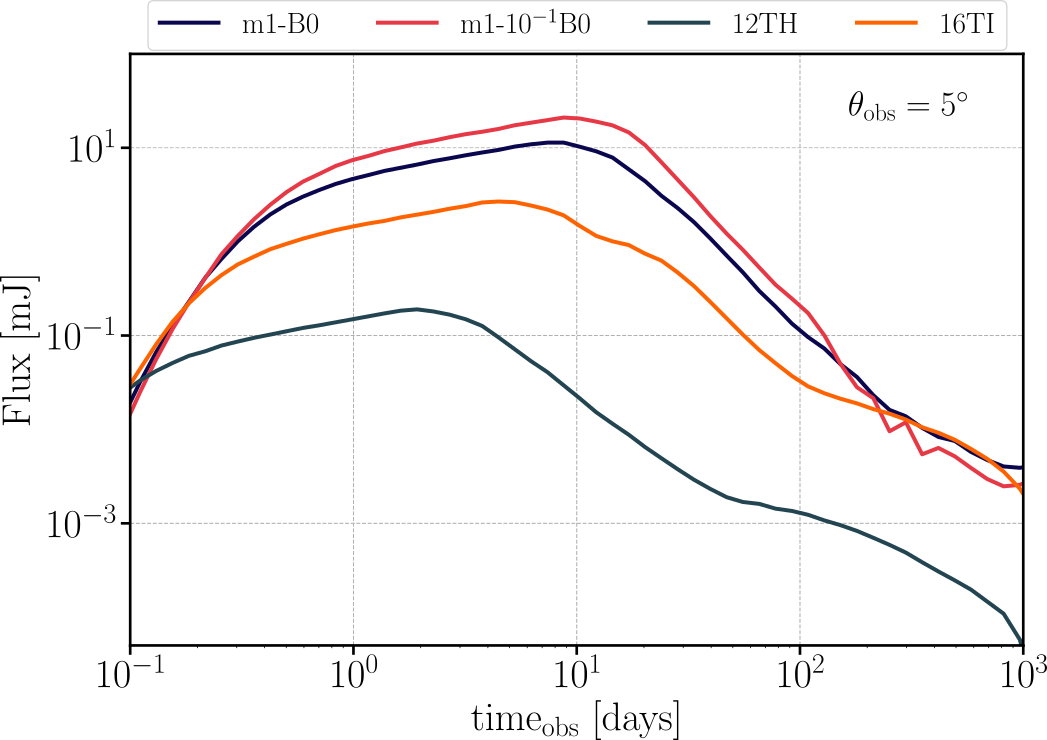}\\
    \includegraphics[width=0.9\linewidth]{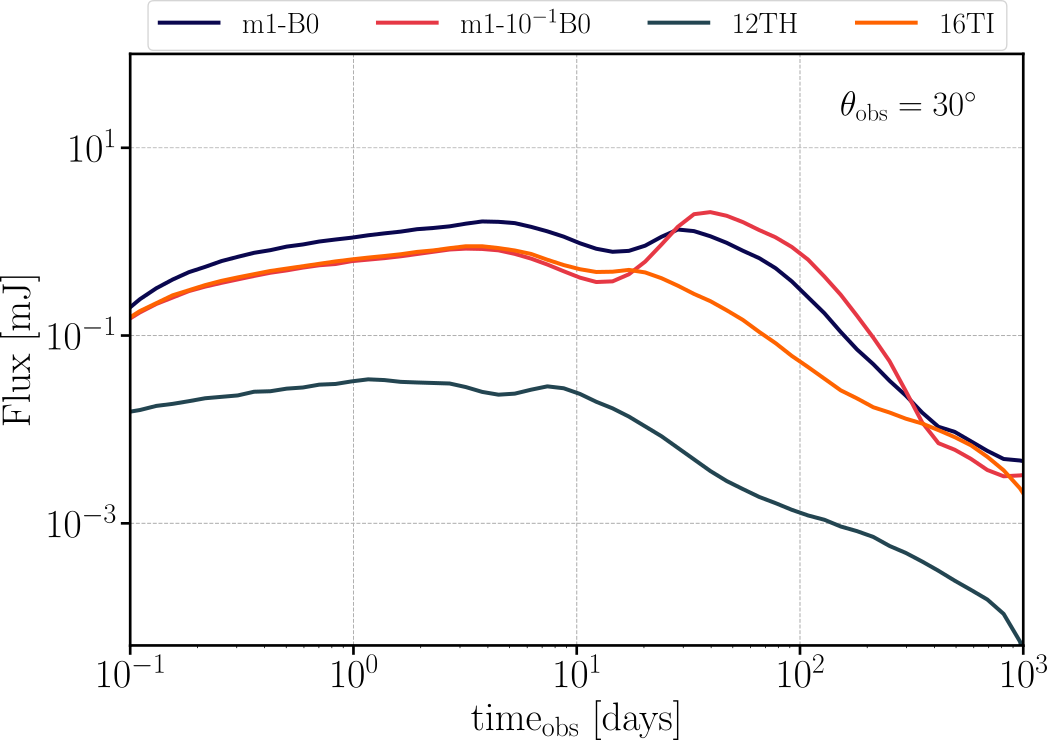}
    \caption{Radio light curves at $3$~GHz for a source located at $40$~Mpc. We assume $\epsilon_e = 0.1$ and $\epsilon_B = 0.01$. The top panel shows the light curve for an observer angle of $\theta_{\rm obs} = 5^\circ$ with respect to the jet axis, and the bottom panel for $\theta_{\rm obs} = 30^\circ$.}
    \label{fig:light_curves}
\end{figure}

\subsection{Jets launched from the BH horizon vs. launching at intermediate distances}

In the literature, the propagation of GRB jets from collapsars has been investigated through hydrodynamical simulations, where the jet is not self-consistently launched from the stellar core. Instead, it is commonly assumed that the jet has already emerged from the iron core, and its injection is implemented through strong jump conditions imposed at the core surface \citep[e.g.,][]{Begelman1989, Morsony2007, lopezcamara2013, lopezcamara2016, Hamidani2017, decolle18b, Matsumoto2019, harrison18, Gottlieb2020_hydro_jets, Suzuki2022, Pais2022, urrutia22_3D}. These studies explore how the choice of initial jet parameters, such as the opening angle $\theta_j$, luminosity $L_j$, injection time $t_j$, and specific enthalpy $h_j$, affects jet propagation.

These parameters collectively control the energy balance between the jet and the stellar envelope, characterised by the ratio $h_{\rm j}\Gamma_{\rm j}\rho_{\rm j}/\rho_{\rm envelope}$ \citep[e.g.,][]{Matsumoto2019}. This ratio governs whether the jet behaves as cold (kinetically dominated) or hot (thermally dominated), with significant implications for cocoon formation and expansion.

A critical point of comparison between hydrodynamical jets and those launched magnetically from a central engine is identifying the dominant energy component after the jet traverses the iron core. The energy balance at this interface provides an indicator for relating the properties of jets imposed at the iron core surface to those self-consistently launched from the BH horizon. We argue that the dominant energy form at $r \sim 10^8$~cm is the key criterion for jet-launching approaches. Thus, a detailed understanding of energy conversion processes on these scales is essential to validate thermal or kinetic jets at the launching point at such distances. Our results (Figure~\ref{fig:fraction_and_radial_energy}) suggest  that conversion from magnetic to thermal (or kinetic) energy at the scale of the iron core may be possible. Consequently, thermal and kinetically dominated jets (see PD and KD models in \citet{urrutia22_3D}) could be self-consistent with our collapsar scenario.

Another key point lies in the influence of the progenitor’s density profile on the jet opening angle \citep[e.g.,][]{Bromberg2011,Hamidani2021-expanding}. The local density determines whether the early jet morphology, as shaped near the central engine, is re-collimated when the jet propagates outwards. Our results show that ongoing accretion leads to time-dependent changes in the density profile, which in turn modify the jet opening angle in the inner part of the progenitor. The jet is collimated as has been previously predicted \citep[e.g.,][]{Hamidani2021-expanding}, however, our outcomes show that it is not a universal result (See Table~\ref{tab:final_outcomes}).

\subsubsection{Implications to the prompt emission}

The GRB prompt spectrum is analysed to determine its peak energy and hardness, which are phenomenologically related to the progenitor type \citep[e.g.,][]{kumar15,Salafiaetal2016}. For this study, our discussion is restricted to GRBs originating from collapsars. The maximum energy for long GRBs have been reported close to $E_{\rm iso}\sim 10^{54}$~erg \citep{Atteia2017,Atteia2025}, our results are bellow to such limit. 

The gamma-ray light curve during prompt emission provides information about the duration and variability in the jet and the maximum duration. Following the Eq.(15) in \citet{gottlieb2022a} The duration of our central engine could be $t_{\rm ce}\sim 100$~s.

The variability is related to activity within the central engine, as instabilities in the accretion disk may lead to intermittent outflows \citep[e.g.,][]{janiuk_variability2021}. Additionally, variability is influenced by turbulence within the jet itself as it penetrates the progenitor star. In our simulations, we observe energy dissipation during the interaction of the jet with the progenitor environment. The variability observed close to the BH horizon could be modified during such jet interaction if the kinetic energy is not dominant, and no periodicity is observed. Then, our jet variability results from a combination of central engine activity and the jet interaction with the progenitor.

The time dependence of the luminosity $L(t)$ has been discussed \citep[e.g.,][]{bobsnjakBarniolPeerLuminosity2022}, revealing how much energy is extracted from the central engine. Analysing the slope of luminosity provides insights into central engine activity. Cases with constant spin establish an upper limit, while changes in spin rates can indicate evolving conditions within the central engine in the first seconds of evolution \citep{Jacquemin-Ide2024}. In our results, the evolution of luminosity $L(t)$ depends mainly on the structure of the progenitor and the accretion (see, for example, Figure~\ref{fig:density_rad} and \ref{fig:ceAct}). The luminosity decreases slowly, and there are no important differences with respect to simulations with spin evolution \citep[e.g.,][]{Shibata2025arXiv}.

\subsubsection{Final jet structure and afterglow emission}\label{sec:afterglow}

We analyse the jet and cocoon structure through the angular energy distribution $E(\theta)$ for material with Lorentz factor $\Gamma > 2$. We found that the progenitor profile is the dominant factor shaping this structure, in particular the distribution of the energy components, as illustrated in Figure~\ref{fig:energy_final_structurse}.

Based on the jet and cocoon structure, we estimate the afterglow emission for light curves using the synchrotron standard model for GRBs \citep{Sari1998,GranotSari2002}. Given an energy $E(\theta)$ and $\Gamma_\infty(\theta)$, such emission is computed by considering the contribution of each angular bin evolves independently as a piece of the shock front until the deceleration radius $R_d\approx(E(\theta)/\rho_0\Gamma_\infty^2(\theta))^{1/3}$. Once the jet and cocoon decelerates at very large distances $d\gtrsim 10^{16}$~cm \citep[e.g.,][]{Ramirez-Ruiz-Mcfadyen_2010}, the deceleration follows \citet{BMK76}
self-similar solution. 

In Figure~\ref{fig:light_curves}, we show the afterglow light curves for two observation angles, respectively, at $\theta_{\rm obs}=5^\circ$ and $\theta_{\rm obs}=30^\circ$. We assume the post-shock thermal energy resides in the accelerated electron population with a fraction of $\epsilon_e=0.1$, while the fraction of magnetic energy at the post-shock region is $\epsilon_e=0.01$. We set a radio frequency of $3$~GHz, an observer's distance of $40$~Mpc. A detailed estimation of such light curves is described in \citep{decolle12}, and afterglow estimation by the final structure of the jet performed by simulations in \citet{Urrutia2021ShortGRBS}.

Figure~\ref{fig:energy_final_structurse} highlights differences in the final jet structure between model m1-$10^{-1}$B0 ($dE/d\Omega \propto \theta^{-2.2}$) and model m16TI ($dE/d\Omega \propto \theta^{-1.3}$), due to the jet-launching and propagation inside different progenitors. Since the afterglow emission is sensitive to the jet and cocoon structure, Figure~\ref{fig:light_curves} illustrates how these differences may manifest at late times and large scales.

However, our modelling has two main limitations. First, although the integration time is short, it is sufficient to capture the breakout and the cocoon engulfing the progenitor. At this point, the jet and cocoon structure is nearly final. As shown in \citet{Urrutia2021ShortGRBS}, the duration of the central engine only influences the normalization of the total energy, and the angular structure of the jet and cocoon is partially preserved. Then, we do not expect significant changes in the angular structure beyond our integration times listed in Table~\ref{tab:models_performed}.

The second limitation is that our light curves were estimated from an analytical blast-wave propagating at large scales, but it does not solve the lateral expansion. Large-scale afterglow predictions have been made through post-processed simulations \citep{vanEerten2010,decolle12,Gill2019,MedinaCovarrubias2023,Govreen-SegalNakar2024,Dreas2026}, however, these do not include self-consistently launched jets.

\subsection{Resolution effects}\label{sec:resolution}

\begin{figure}[h!]
    \centering
    \includegraphics[width=1.0\linewidth]{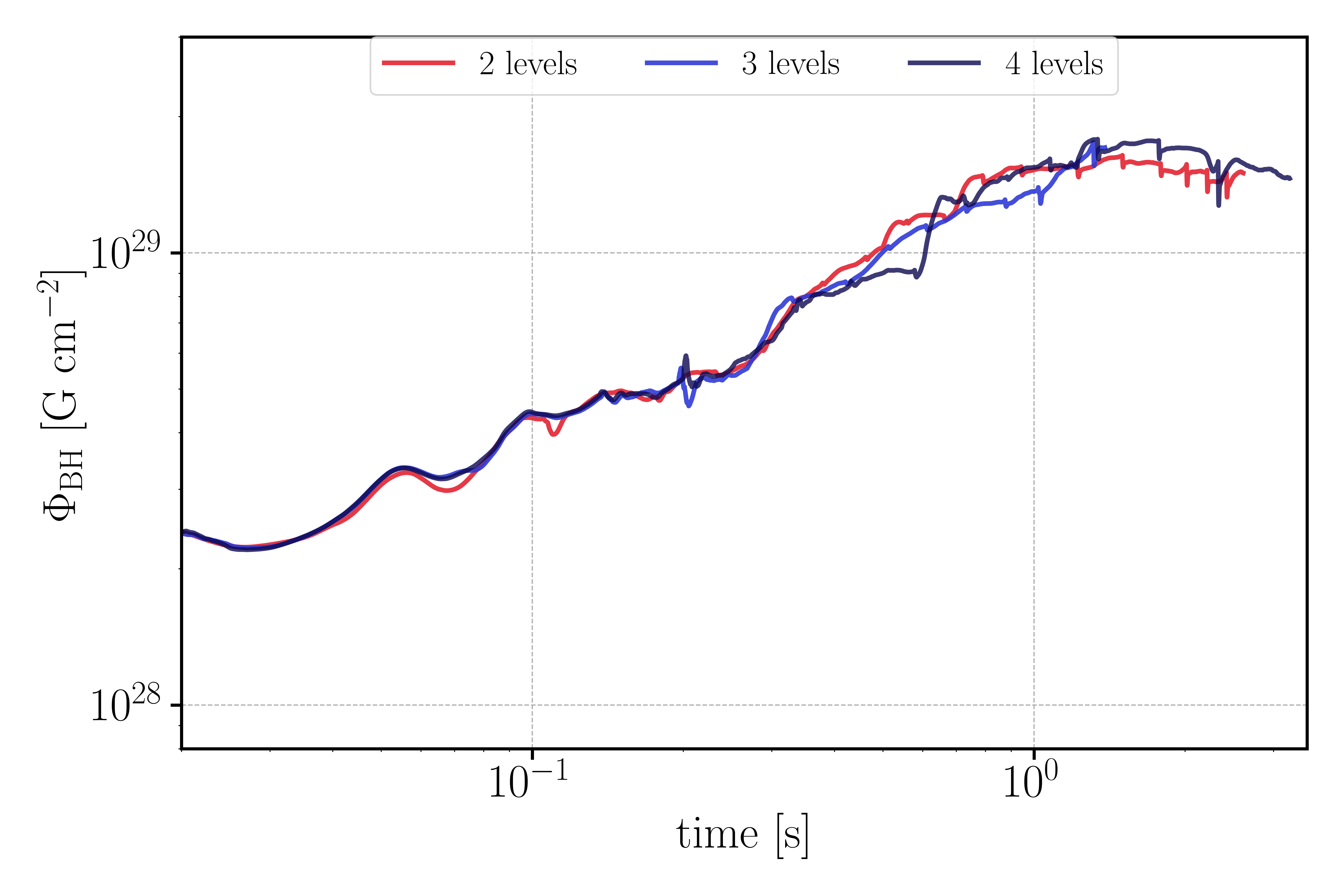}
    \includegraphics[width=1.0\linewidth]{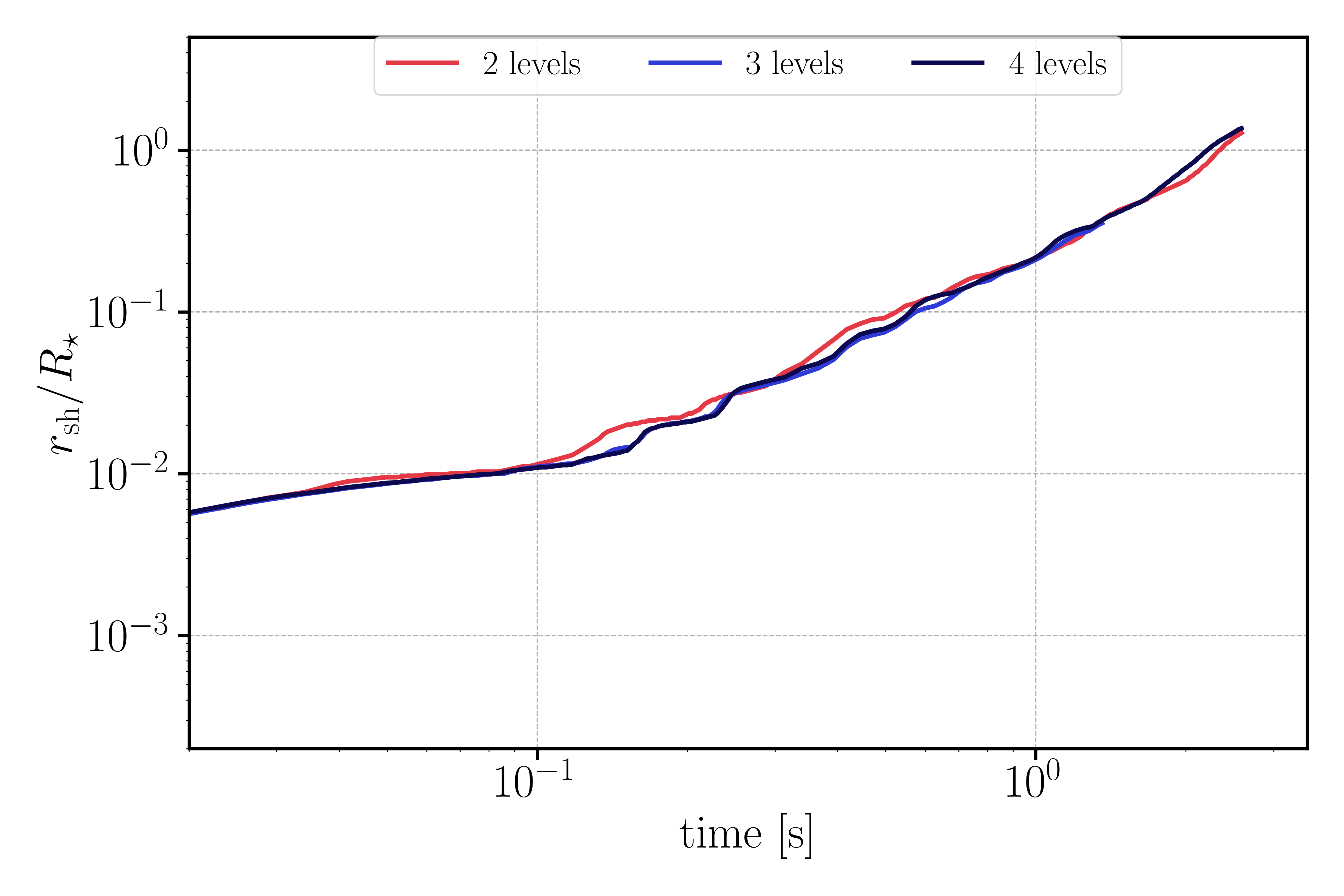}
    \caption{Convergence test for model m1-B0 using three different levels of refinement ($n_l=$~2, 3, and 4). The upper panel displays the time evolution of the magnetic flux. The lower panel shows the evolution of the shock front position. The resolution does not affect the evolution of the magnetic flux and the jet propagation.}
    \label{fig:convergence}
\end{figure}

As a resolution test, we perform model m1-B0 with three different levels of refinement: $n_l=$ 2, 3, and 4. In Figure~\ref{fig:convergence}, we present the convergence test. The upper panel shows the magnetic flux (eq.~\eqref{eqn:magnetic_flux}), which is essential for driving Poynting-flux-dominated jets. In addition, since we track the propagation and breakout of the jet, the lower panel shows the evolution of the shock front, which is not affected by the resolution. We choose $n_l=3$ in our set of simulations (Table~\ref{tab:models_performed}), as this configuration follows a trend similar to that of the simulation with four levels.  Then, our maximum angular resolution is~$\sim 0.7^{\circ}$, which is comparable to the angular resolution in the $\theta$ direction adopted in similar studies \citep[e.g.,][]{gottlieb2022a}.

The adopted resolution is sufficient to resolve the jet core. For example, \citet{Lloyd-Ronning2020} reports narrow jets from an observational sample, have opening angles of $\theta_j \sim 6^{\circ}$. We assume this angle as the minimum representative of standard GRBs. However, our setup could be limited to resolve unusually narrow jets, such as GRB~221009A (``THE~BOAT''), which has an opening angle in a range of $\theta_j \sim 1.5^{\circ}-3.2^{\circ}$ \citep{Burns2023,OConnor2023,Axelsson2025}.

Finally, we don't observe the development of plug instability as a consequence of the angular resolution or the axisymmetric computational box. A comparison of jet propagation with and without plug instability is presented in Appendix~A of \citet{Urrutia2021ShortGRBS}.

\section{Conclusions}\label{sec:conclusions}

We performed a set of axisymmetric GRMHD simulations that follow the entire propagation of collapsar jets from the black-hole horizon to the jet’s emergence at the stellar surface.  We studied the impact of the progenitor density profile, the strength of the magnetic field, and the field geometry (pure dipole versus hybrid) on jet formation and subsequent explosive activity on Long GRBs.  The main conclusions are:

\begin{itemize}
  \item[-] Jets are launched only when a dipolar field with peak strength $B_0\gtrsim10^{12}$~G. It produces a magnetic flux $\Phi_{\rm BH}\gtrsim10^{25}$~G~cm$^2$.  Such models quickly reach a magnetically arrested–disk state with $\phi_{\rm MAD}\gtrsim15$ and produce jet luminosities $L_{j}> \dot M c^{2}$, implying efficient powered energy extraction via the Blandford–Znajek process. Weaker or non‐magnetised configurations fail to collimate and never break out of the star.
  
  \item[-] Strongly magnetised models develop a narrow, highly magnetised core ($\sigma\gtrsim1$) surrounded by moderately magnetised wings.  Increasing $B_0$ tightens collimation and sharpens the core–wing contrast, whereas hybrid field geometries yield a quasi‐cylindrical outflow described as a \emph{failed} jet. At jet's breakout, the magnetisation drops and kinetic energy dominates, setting the stage for the afterglow.

  \item[-] For identical magnetic setups, the Wolf–Rayet models 12TH and 16TI produce stronger core–wing mixing and larger terminal opening angles. The jet collimation is affected after the breakout time for jets launched from MAD state central engines, which impacts previous predictions from hydrodynamical simulations and semi-analytical models. Our results suggest that after breakout, the jet core has two components: magnetic and kinetic. The magnetic component fishes collimated and the kinetic expanded.

  \item[-] The non-magnetized accretion (model m1-zero) never excavates a low-density funnel. In addition, we observe an expanding bubble whose shock front is sustained by winds ejected from the disk until $t\sim 1$~s. After this time, an additional energy source (either from magnetic flux or neutrino heating) is required to sustain the shock front for longer times and produce breakout from the star.

  \item[-] The successful jets carry $L_{j}\sim10^{49}$-$10^{53}\,$erg\,s$^{-1}$ and decelerate into a structured outflow whose angular energy profile $E(\theta)$ reproduces the shallow cores and steep wings inferred from GRB afterglow modelling.  Synthetic 3\,GHz light curves show bright, rapidly rising on‐axis emission and late–peaking for off‐axis observer.
\end{itemize}

Our calculations neglect neutrino transport, assume a fixed black–hole spin ($a\sim0.9$) and enforce axisymmetry. Full 3-D simulations including neutrino heating and spin evolution will be required to track turbulence, jet wobbling, and long‐term engine variability self‐consistently.  Nonetheless, the present results indicate that (i) the progenitor’s inner density gradient and (ii) the large‐scale dipolar field impact the dynamics and structure of a relativistic-magnetically dominated jet.

\bigskip

\section*{Acknowledgements}
This work was supported by the grants 2019/35/B/ST9/04000 and 2023/50/A/ST9/00527 from Polish National Science Center. We gratefully acknowledge Polish high-performance computing infrastructure PLGrid (HPC Center: ACK Cyfronet AGH) for providing computer facilities and support within computational grant no. PLG/2025/018086. This work was also supported by the Center for Research and Development in Mathematics and Applications (CIDMA) through the
Portuguese Foundation for Science and Technology (FCT), references UIDB/04106/2020, UIDP/04106/2020 (\url{https://doi.org/10.54499/UIDB/04106/2020} and \url{https://doi.org/10.54499/UIDP/04106/2020}). HO acknowledges support from the projects PTDC/FIS-AST/3041/2020, CERN/FIS-PAR/0024/2021 and 2022.04560.PTDC, the European Union’s Horizon 2020 research and innovation (RISE) programme H2020-MSCA-RISE-2017 Grant No. FunFiCO-777740 and by the European Horizon Europe staff exchange (SE) programme HORIZON-MSCA-2021-SE-01 Grant No. NewFunFiCO-10108625. We acknowledge helpful discussions with Ore Gottlieb, Miguel Aloy, Fabio De Colle, and we thank Ariadna Murguia-Berthier, Leonardo García-García, Ludovica Crosato-Menegazzi, Rosa Becerra and David Aguilera-Dena for useful comments. 




\bibliographystyle{elsarticle-harv} 
\bibliography{example}






\end{document}